\documentclass[aps,twocolumn,showpacs]{revtex4-1}
\usepackage{amsfonts}
\usepackage{amsthm}
\usepackage{amsmath}
\usepackage{amsfonts}
\usepackage{latexsym}
\usepackage{amssymb}
\usepackage{xcolor}
\usepackage[ansinew]{inputenc}
\usepackage{graphicx,fancyhdr}
\usepackage{lastpage}
\usepackage{epstopdf}
\usepackage{graphics}
\usepackage{latexsym}
\usepackage{amsmath}
\usepackage{dsfont}
\def\beqn{\begin{eqnarray}}
\def\eeqn{\end{eqnarray}}
\def\beqa{\begin{eqnarray}}
\def\eeqa{\end{eqnarray}}
\def\non{\nonumber}

\def\hw{\hbar \omega}
\def\hw4{ \frac {\hbar \omega}{4}}

\begin{document}
\title{Excepional Points from Hamiltonians of hybrid physical systems: Squeezing and anti-Squeezing.}
\vskip2cm
\author{Ram\'\i rez R. $^{a)}$ \footnote{e-mail: romina@mate.unlp.edu.ar}}
\author{Reboiro M. $^{b)}$ \footnote{e-mail: reboiro@fisica.unlp.edu.ar}}
\author{Tielas D. $^{c)}$ \footnote{e-mail:diegotielas@gmail.com}}
\affiliation{{\small\it $^{a)}$Department of Mathematics, National University of La Plata}
{\small \it La Plata,Argentina}}
\affiliation{{\small\it $^{b)}$IFLP, CONICET-Department of Physics, National University of La Plata}
{\small \it La Plata, Argentina}}
\affiliation{{\small\it $^{c)}$IFLP, CONICET-Faculty of engineering, National University of La Plata}
{\small \it La Plata, Argentina}}
\date{\today}

\begin{abstract}
We study the appearance of Exceptional Points in a hybrid system composed of a superconducting flux-qubit and an ensemble of nitrogen-vacancy colour centres in diamond. We discuss the possibility of controlling the generation of Exceptional Points, by the analysis of the model space parameters. One of the characteristic features of the presence of Exceptional Points, it is the departure from the exponential decay behaviour of the observables as a function of time. We study the time evolution of different initial states, in the presence of the hybrid system, by computing the reduced density matrix of each subsystem. We present the results we have obtained for the steady behaviour of different observables. We analyse the appearance of Squeezed Spin States and of anti-Squeezed Spin States.
\end{abstract}

\maketitle

keywords: non-hermitian Hamiltonians, Exceptional Points, Squeezing, anti-Squeezing, Schr\"odinger-cat states.

\section{Introduction}

Quantum information processing \cite{qip0,qip} can involve the interplay of atoms in optical traps \cite{traps,traps1}, superconducting circuits \cite{super,qsupremacy}, nuclear spins \cite{nucsp} and defects in a crystal lattice \cite{crystals}, among others physical components. Usually, the system to be manipulated interacts with its environment, and this interaction degrades its quantum dynamics \cite{deco0}. It can be distinguished two very different cases of coupling with the environment. In the first case, the environment consists of a continuum of scattering wave functions which can mediate the escape of particles from the localized system, i.e. unstable states in a nuclei \cite{feschbach,rotter}. In the other case, the environment is provided by the states of a macroscopic reservoir, and the strength of coupling depends on the overlap between states of the localized system and states of the reservoir, i.e. quantum-dot transport\cite{ferry,ferry1,longhi2}.

In the last years, in different experiments \cite{patowski,patowski1,persson,ferry1,new1}, it has been reported conditions under which, due to the coupling with the environment,  the transition between two different dynamical regimes can be observed, an oscillating regime and an overdamped regime. This transition is not a smooth crossover but it has the characteristics of a critical phenomena. It has received the name of quantum Dynamical Phase Transition (DPT) \cite{patowski}.
One of the earliest works to explicitly invoke the notion of a DPT involved studies of spin swapping in atomic systems \cite{patowski}.
In these experiments, the authors of \cite{patowski} have studied Rabi oscillations due to spin flips,
 and have observed transition to strongly damped motion by increasing the coupling to an environment
formed by a spin bath. Evidence of DPTs is present also, in an earlier work on resonance trapping in microwave cavities \cite{persson},
and on einselection \cite{ferry1} in mesoscopic quantum dots. In both systems, the focus was on understanding how the states of a quantized cavity are affected by coupling them to an external environment. The phenomenon shown in both cases
was that certain eigenstates actually became narrower when the environmental coupling was increased over a specific range. The accompanying short-lived eigenstates appear as a background with which the long-lived ones interfere, a behaviour that has been observed experimentally \cite{persson}. The long-lived states may even become discrete, forming so-called bound states in the continuum \cite{rotter0}. It is interesting to note that the presence of long-lived states is a problem that dates back to the earliest days of quantum mechanics \cite{vonNeumann}.

From the theoretical point of view, the dynamics of open physical systems can be described in the framework of non-hermitian Hamiltonians \cite{rotter,feschbach,faisal,open1,open2,gadella0,gadella,savannah0,savannah1}. In the study of a parametric family of non-hermitian Hamiltonians, it is usual to observe regions with different symmetry. These zones are determined by the properties of the spectrum. The transition occurs when two or more eigenvalues coalesce into one and the corresponding eigenvectors become parallel. Such degeneracies are called Exceptional Points (EPs). Non-hermitian Hamiltonians arise naturally within the formalism of Feshbach \cite{rotter,feschbach,faisal}. In this approach, the space of the system under study is broken into two subspaces, the subspace corresponding to the localized system, and the subspace related to the extended environment. 
The solution of problem in the whole function space (localized system embedded in a well-defined environment), which is described by a hermitian Hamiltonian operator, can be projected in the interior of the localized
part of the system, by a set of eigenfunctions of an
effective non-hermitian Hamiltonian. Within Feschbach's formalism, the corresponding matrix elements describing the coupling
that develops between the different states of the localized system are typically
complex \cite{rotter,feschbach,faisal}.

Among non-hermitian Hamiltonians, pseudo-hermitian Hamiltonians play a central role. The formal beginning of this subject was due to Bender and Boettcher \cite{bender,benderrep} in 1998. Since then, they have proven to be very useful in the understanding of physical problems with manifest Parity-Time Reversal ($\mathcal{PT}$) symmetry, i.e. microwave cavities \cite{8}, atomic diffusion \cite{12}, electronic circuits \cite{15}, optical waveguide arrays \cite{joglekar}, quantum critical phenomena \cite{ueda,mikhail,mikhail1}.

Recently, it has been carried out an experiment on a superconducting qubit with dissipation close to its EP \cite{scexp}. Using quantum tomography the authors of \cite{scexp} have studied the steady-state of the qubit system evolving under non-hermitian dynamics in the vicinity of the EP. The results indicate that the dissipation of the system stabilizes the qubit to non-trivial steady-states for different drive and detuning parameters.
In the same line, the Authors of \cite{nv2019} have shown that microwave driven Nitrogen-Vacancy(NV) colour centres can function as robust mode switches in the vicinity of an EP. These works point out the need to explore and harness EP degeneracies for enhanced sensing and quantum information processing.

In this direction, hybrid devices \cite{hybrid} are promising candidates for the realization of quantum processing. In the present work, we investigate the presence of EPs in a hybrid system consisting of a superconducting flux qubit (SFQ) and an ensemble of NV$^-$-colour-centres in diamond \cite{zhu,zhubis,marco,nv-qb-1,nosap17,noswig1,noswig2}. We shall show that robust steady-states in the vicinity of EPs can be constructed by an adequate selection of the initial state. In particular, we shall show that, due to the non-hermitian dynamic of the system, for certain values of the parameters of the model space, the steady-state is maximally anti-squeezed \cite{antisq,antisq1} and behaves as a Schr\"odinger-cat state (SCS) \cite{cat0,cat1,cat2}.

The work is organized as follows. The details of the formalism are given in Section \ref{formalism}. We describe the  physical hybrid system under consideration and we present the time evolution of a give initial state within the formalism of Green Matrix.
The results of the calculations are presented and discussed in Section \ref{results}. We evaluate the appearance of EPs as a  function of the space of parameters of the model. We study the time evolution of different initial states and we discuss the possibility of generating steady Schr\"odinger-cat states. Our conclusions are drawn in Section \ref{conclusions}.

\section{Formalism}\label{formalism}

We shall consider the Hamiltonian of an hybrid system, composed by a superconducting flux-
qubit coupled to an ensemble  of NV$^{-}$ colour centres in diamond \cite{zhu,zhubis,marco,nv-int-2,nv-1,nv-int-1,nv-int-new,hybrid-10,hybrid-11,nosap17},

\beqa
H_{hybrid} & = & H_{Fq}+H_{S}+H_{int-qs}.
\label{h0}
\eeqa

The term $H_{Fq}$ of Eq. (\ref{h0}) is the Hamiltonian of the superconducting  flux-qubit. In the basis of clockwise and
anticlockwise qubit-persistent-currents \cite{zhu,hummer,orlando,plourde,nos-jun,nosap17}, it reads
\beqa
H_{Fq} & = & \frac 12 ~\left( ~\Delta ~s_{x} + ~\epsilon s_{z} \right),
\label{hf}
\eeqa
where $\{s_{x},~s_{y},~s_{z}\}$ are Pauli spin-1/2  operators.
The parameter $\epsilon$ is the energy bias,
$\epsilon=2 I_{p}(\Phi_{ex}-3\Phi_0/2)$, $I_{p}$ is the persistent
current in a qubit, $\Phi_{ex}$ is the external flux threading the
qubit loop, $\Phi_0=1/(2 e)$ is the flux-quantum and $\Delta$ is the
tunnel splitting.

The term $H_S$ of Eq.(\ref{h0}) models the ensemble of NV$^{-}$ colour centres in diamond. It reads

\beqa H_{S}  & = & D ~S^2_z + E ~ (S_x^2-S_y^2). \label{hs} \eeqa

The operators
$\{S_x,~S_y,~S_z\}$ are Pauli spin-operators, components of the total spin, ${\bf S}$, of the NV$^{-}$ ensemble. Being D the zero-field splitting (2.878 GHz) and E the strain-induced splitting \cite{marco}.
This sector of the Hamiltonian consists of a one-twist term
\cite{kitagawa}, $D ~S^2_z$, responsible for the squeezing
pattern, and a Lipkin-type interaction $ E ~ (S_x^2-S_y^2)$  \cite{ring,lipkin,noschains}.

The coherence of the ensemble of NV$^-$-colour-centres highly depends on the contents
of paramagnetic impurities in diamond \cite{zhu,nv-ct1,nv-ct2,nv-ct3,linew1,linew2}. The main magnetic impurities are neutral nitrogen atoms, $P1$ centres. The contribution of $P1$ centres to the decoherence of the NV ensemble has been thoroughly studied \cite{linew1,linew2}. In particular, it was shown that the coherence time of the NV ensemble depends on the concentration of $P1$ centres. Due to the presence of $P1$ centres in the the ensemble, we shall replace the coupling interaction of the NV centres with the SFQ \cite{zhu}, $(g/2)~s_{z}~S_x $, by an asymmetric interaction of the form

\beqa
H_{int-qs} & = & \frac 14~ ~ g  ~s_{z} (S_+ + \alpha S_-) \nonumber \\
 & = & \frac 14~ ~ g  ~s_z \left((1+\alpha) S_x + {\bf i} (1-\alpha) S_y \right).
\label{hint1}
\eeqa
A similar approach have been reported \cite{asymmetry,asymmetry1}.

The gap-tunable flux-qubit Hamiltonian, $H_{Fq}$ of Eq. (\ref{hf}), is diagonalized by
the transformation

\beqa \left (
\begin{array}{c}
s_{z} \\
s_{x} \\
s_{y}
\end{array}
\right) = \left (
\begin{array}{ccc}
 ~ \cos \beta  & \sin \beta  & 0 \\
-\sin \beta  & \cos \beta  & 0 \\
0 & 0 & ~1 \\
\end{array}
\right) \left (
\begin{array}{c}
\sigma_{z} \\
\sigma_{x} \\
\sigma_{y}
\end{array}
\right).\non \\
\label{transf}
\eeqa

Both set of operators, $ \{\sigma_{x}, \sigma_{y}, \sigma_{z}\} $ and
$\{s_{x},~s_{y},~s_{z}\}$, obey the $su(2)$ algebra. The parameter $\alpha$ of the transformation is related to parameters of the qubit flux by
$\cos \beta= \epsilon /E_{qb }$,
$\sin \beta =-\Delta /E_{qb }$, and
$E_{qb }=\sqrt{\epsilon ^2+\Delta ^2}$.

In terms of the new operators $\sigma_i$ $(i=~x,~y,~z)$, $H_{Fq}$ and $H_{int-qs}$ can be written as

\beqa
H_{1}   & = &  ~ \frac 12 E_{qb } \sigma_{z }, \nonumber \\
H_{2} & = & \frac 14~ g   \left(~\cos \beta ~\sigma_{z } + \sin \beta  ~\sigma_{x } \right) (S_+ + \alpha S_-). \nonumber \\
\label{hijc}
\eeqa



Finally, the Hamiltonian of the hybrid system, $H$, is given by the sum of the terms (\ref{hs}) and (\ref{hijc}), by

\beqa
H= H_{1}+H_{2}+H_S.
\label{hami}
\eeqa
In general, this Hamiltonian is a non-hermitian operator, except if $\alpha=~1$. We can aim to diagonalize $H$ in the product basis

\beqa
| k_{qb}, N_S, k_S \rangle & = & | k_{qb} \rangle \otimes |N_S, k_S \rangle, \non \\
| N_{S}, k_{S} \rangle & = & {\cal N}_{S}~S_+^{k_{S}} |0 \rangle_{S},\non \\
| k_{qb} \rangle & = & {\cal N}_{qb}~\sigma_{+}^{k_{qb}} |0 \rangle_{qb},\non \\
\label{base}
\eeqa
where $S_\pm = S_x \pm i S_y$. The label $k_S$ can run
from $0$ to the number of spins (electrons) of the system, N. Similarly, $\sigma_{\pm,j}=\sigma_{x,j} \pm i \sigma_{y,j}$, and $k_{qb,j}=0,~1$. We denote with
$\{k_{qb,j}\}$ each of arrays of $N_{qb}$ superconducting  qubits.
The quantities $\cal N$ are normalization factors.

The Hamiltonian of Eq.(\ref{h0}) is invariant under parity symmetry but not under time-reversal symmetry, consequently, it is not invariant under ${\cal PT}$ symmetry. This can be seen from the transformation properties of the operators \cite{mottelson}.
The linear parity operator ${\mathcal P}$ performs spatial reflection, so that the position, the momentum and the spin transform as
${\bf r} \rightarrow - {\bf r}$,
${\bf p} \rightarrow - {\bf p}$, and
${\bf S} \rightarrow   {\bf S}$, respectively. Whereas,
time-reversal operation can be represented by an anti-unitary
operator ${\mathcal T}=U~ K$, being $U$  an unitary operator and $K$ the complex conjugation operator \cite{mottelson}. Under time reversal, we have
${\bf r} \rightarrow  {\bf r} $,
${\bf p} \rightarrow  {\bf p} $,
${\bf S} \rightarrow -{\bf S} $ and
${\rm \bf i} \rightarrow -{\rm \bf i}$.
The ${\cal T}$-symmetry is broken by the first term of $H_1$
of Eq. (\ref{hijc}). Notice that the states of the basis of Eq.(\ref{base}), can be classified according to its transformation under parity

\beqn
& & {\cal P}\mid k_{qb}, N_S, k_S \rangle = \nonumber \\
& & ~~~~~~~~~~~(-1)^{k_S+k_{qb}}\mid k_{qb}, N_S, k_S \rangle.
\eeqn

Nevertheless, $H$ and $H^{\dagger}$ are quasi-hermitian operators. This can be proved straightforward by observing that $H^{\dagger}=H^T$, so that they are isospectral Hamiltonians

\beqn
H & = & {\widetilde P} J {\widetilde P}^{-1} \nonumber \\
H^T & = & {\overline P} J {\overline P}^{-1}. \nonumber \\
\eeqn
In general, $J$ is a Jordan matrix, while ${\widetilde P}$ and ${\overline P}$ are the matrix of the generalized eigenvectors of $H$ and $H^T$, respectively. Finally, the symmetry operator ${\cal S}$, such that
$H={\cal S} H^T {\cal S}^{-1}$,
is given by ${\cal S} ={\widetilde P} {\overline P}^{-1}$ .

Depending on the values of the parameters in the space $\{\epsilon ,~ \Delta,~D,~E,~g,~\alpha \}$, the spectrum of the Hamiltonian can consist of real eigenenergies or  complex-conjugate pairs eigenenergies. Also, for particular values of the parameters, exceptional points, the coalescence of two or more eigenvalues is accompanied by the coalescence of the corresponding eigenvectors. Depending on the characteristics of the spectrum of the Hamiltonian of Eq. (\ref{h0}), the time evolution of an initially prepared state will display different behaviours. In the regime of real spectrum, the mean value of the observables will display a patron of collapses and revivals. While, in the regimen of complex-conjugate pair spectrum, the observables will show the behaviour of systems with gain-loss balance \cite{gainloss}. Besides, in the case of exceptional points, the time evolution of the mean values of observable will present departures from the usual exponential decay behaviour. We shall vary the values of $\{~g/E,~\alpha \}$ to control the generation of which EPs.

\subsection{Time Evolution}

The dynamics generated by the Hamiltonian of Eq. (\ref{hami}) can be captured by introducing the non-Hermitian prescriptions given in \cite{jmp19,optimal}.
Alternatively, we can use the Green operator formalism \cite{green,patowski1}.
Let us briefly review the essentials of it.

Associated to the Hamiltonian $H$, we can introduce the Green operator, ${\bf G}$. It reads

\beqn
{\bf G} (\omega)& = & (\omega {\bf I}-H)^{-1}.
\eeqn
The time evolution of the system is related to the Fourier Transform of the Green operator, $\bf{\mathcal{F}}$.
To compute $\bf \mathcal{F}$, we shall define the retarded and the advanced Green operators, $G^-(\omega)$ and  $G^+(\omega)$ respectively, as

\beqn
{\bf G}^\mp (\omega)& = & ((\omega \pm {\bf i}\eta) {\bf I}-H)^{-1}.
\eeqn
Its Fourier Transform, ${\mathcal F}^{\mp}(t)$, is given by

\beqn
{\mathcal F}^\mp(t) & = & \int_{-\infty}^{+\infty} \frac {\rm d \omega}{2 \pi}~  {\bf G}^\mp(\omega) {\rm e}^{-{\bf i} \omega t}.
\eeqn
Given ${ \mathcal F}(t)={\bf i } { \mathcal F}^+(t)-{\bf i } {\mathcal F}^-(t)$, we can introduce the transition matrix ${\rm P}$ as

\beqn
{\rm P}(t)= {\cal N}^2(t)~ {\mathcal F}^\dagger(t){\rm \mathcal F}(t),\label{transk}
\eeqn
with ${\cal N}(t)=\left( {\rm Tr}({\mathcal F}^\dagger(t).{\mathcal F}(t))\right)^{-1/2}$.

The transition probabilities, between the  states of Eq. (\ref{base}), are given by the matrix elements of ${\rm P}$.



In the basis of Eq.(\ref{base}), a general initial state can be written as

\beqn
| I \rangle= (c_1,~c_2,~...,~c_n)^T,
\label{ini00}
\eeqn
and it evolves in time as

\beqn
|I(t) \rangle & = &
{\cal N}(t)~ {\cal F}  |I(0) \rangle . \nonumber \\
\eeqn

We are able to evaluate
the mean value of an operator $\widehat{o}$ as a function of time as
\beqn
\langle \widehat{o}(t) \rangle & = & {\langle I (t)| \widehat{o}|I(t) \rangle}.
\eeqn


We shall prepare the initial state of the hybrid system as a direct product of the initial state of the superconducting  qubits and  the NV ensemble initial state
\begin{eqnarray}
|I(0)\rangle =|I \rangle_{qb} \otimes | I\rangle_{NV}. \label{istate}
\end{eqnarray}

For the initial state of the electron ensemble, we shall consider a coherent spin-states of the form

\begin{eqnarray}
| I \rangle_{NV}= {\cal N}_S~{\rm e}^{ z_S S_+ }\mid 0>,\label{zat}
\end{eqnarray}
with $z_{S} =-{\rm e}^{-i \phi_{S}} \tan(\theta_{S}/2)$,
where the angles $(\theta_{S},\phi_{S})$ define the direction
${\bf n}_{S}=(\sin{\theta_{S}} ~\cos{\phi_{S}},\sin{\theta_{S}} ~\sin{\phi_{S}},\cos{\theta_{S}})$, such that
${\bf S} \cdot {\bf n}_{S} |I\rangle_{NV}=-S | I \rangle_{NV}$, with $S={\rm N}/2$ \cite{hecht}.

For the initial state of the superconducting  qubits, we shall consider a particular state of the form $|k_{qb}\rangle $
(Eq.(\ref{base})).

\subsection{Physical Observables}

We shall discuss the effect of non-hermitian term $H_{int-qs}$, Eq. (\ref{hint1}) upon physical observables.

The survival probability, ${\rm p}(t)$, of a given initial state as a function of time is computed as

\beqn
{\rm p(t)} = | \langle I(0)| I(t) \rangle |^2,
\label{prob}
\eeqn
where $| I(0)\rangle$ is the state at which the system has been initially prepared.

Another observable of the system is the squeezing spin parameter, which is related to the reduction of the fluctuations in one of the components of the total spin, bellow the quantum limit, at expenses of the increment of another component of the total spin. There is not a unique way to characterize uncertainty relations of complementary operators \cite{nori,portesi}. We choose to use a squeezing parameter invariant under rotations \cite{luis}. Following the work of Kitagawa and Ueda \cite{kitagawa}, we shall
define a set of orthogonal axes $\{ {\bf n_{x'}}, {\bf n_{y'}},
{\bf n_{z'}} \}$, such that ${\bf n_{z'}}$ is the unitary vector pointing along the direction
of the total spin $<{\bf T }>$.  We shall fix the direction ${\bf n_{x'}}$ looking for
the minimum value of $\Delta^2 T_{x'}$, so that the Heisenberg  Uncertainty Relation reads

\beqn
\Delta^2 T_{x'} ~\Delta^2 T_{x'}~& \ge & ~\frac 14 |\langle T_{z'} \rangle|^2.
\label{hur}
\eeqn
Consequently, the squeezing parameters \cite{kitagawa} are defined as

\begin{eqnarray}
\zeta^2_{x'}  =  \frac {2 \Delta^2 T_{x'} }{|\langle T_{z'} \rangle|},~
\zeta^2_{y'}  =  \frac {2 \Delta^2 T_{y'} }{|\langle T_{z'} \rangle|}.
\label{sqx}
\end{eqnarray}
The state is squeezed in the $x'$-direction if $\zeta^2_{x'}<1$ and $\zeta^2_{y'}>1$.

Also, we shall use the discrete $SU(2)$ Wigner distribution \cite{wigner1,wigner2,wigner3,noswig1,noswig2} to study the time evolution of a given initial state under the action of the Hamiltonian of Eq.(\ref{h0}).

\section{Results and discussion.}\label{results}

We shall consider the case of two NV$^{-}$ colour centres in interaction with a SFQ. As experimentally, the values of $\epsilon $ and $\Delta $ can be controlled independently by using different external magnetic flux \cite{zhu}, we have fixed the value of $\epsilon =0$, so that $E_{qb }=\Delta/2$. For the present case, the model space is divided into two independent blocks, according to the transformation of the vectors of the basis under parity. Ordering the basis as $\{|k_{qb},k_S \rangle\}=\{|0,0\rangle,~|1,1\rangle,~|0,2 \rangle,|1,0\rangle,~|0,1\rangle,~|1,2 \rangle \}$, the Hamiltonian can be written as

\beqn
H= \left (
\begin{array}{rrrrrr}
D-\frac \Delta 4 & E & \frac{g}{2 \sqrt{2}} & 0 & 0 & 0 \\
E & D- \frac \Delta 4 &  \frac{\alpha g}{2 \sqrt{2}} & 0 & 0 & 0 \\
\frac{\alpha g}{2 \sqrt{2}} &\frac{g}{2 \sqrt{2}} & \frac \Delta 4 &  0 & 0 & 0 \\
0 & 0 & 0 & -\frac \Delta 4 & \frac{\alpha g}{2 \sqrt{2}} & \frac{g}{2 \sqrt{2}}\\
0 & 0 & 0 & \frac{g}{2 \sqrt{2}}& D+ \frac \Delta 4 & E\\
0 & 0 & 0 & \frac{\alpha g}{2 \sqrt{2}} & E & D+ \frac \Delta 4 \\
\end{array}
\right )\nonumber \\
\eeqn

In terms of the parameters of the model, $\{\Delta ,~D,~E,~g,~\alpha \}$, the eigenenergies of the Hamiltonian of Eq. (\ref{hami}) are given by

\beqn
E_{1 \pm} & = & \frac 1 2 D+\frac E 6 \left ( d_\pm-2\left(  {\rm e}^{-{\bf i}\theta} \frac{ C_\pm}{  |R_\pm|} +  {\rm e}^{{\bf i}\theta}|R_\pm|\right)\right), \nonumber \\
E_{2 \pm} & = & \frac 1 2 D+ \frac E 6 \left ( d_\pm
+2  \left(
{\rm e}^{-{\bf i}(\frac \pi 3+\theta)} \frac{C_\pm }{ |R_\pm|}+{\rm e}^{{\bf i}(\frac \pi 3+\theta)} |R_\pm|\right) \right),
\nonumber \\
E_{3 \pm} & = & \frac 1 2 D+\frac E 6 \left ( d_\pm
+2  \left(
{\rm e}^{{\bf i}(\frac \pi 3-\theta)} \frac{C_\pm }{ |R_\pm|}+{\rm e}^{-{\bf i}(\frac \pi 3-\theta)} |R_\pm|\right)\right),
\nonumber \\
\eeqn

with

\beqn
d_\pm & = & \left( 1 \pm  \delta \right)\frac D E, \nonumber \\
A_\pm & = & - 27 \gamma^2  (1+\alpha^2)+ d_\pm^3-9 (1-2 \gamma^2 \alpha) d_\pm, \nonumber \\
C_\pm & = & 3 (1+4 \gamma^2 \alpha)+ d_\pm^2, \nonumber \\
B_\pm & = & A_\pm^2-C_\pm^3, \nonumber \\
R_\pm & = & (A_\pm+ \sqrt{B_\pm})^{1/3}= |R_\pm|{\rm e}^{{\bf i} \theta},
\eeqn
given in terms of the adimensional parameters $\delta=\Delta/(2 D)$ and $\gamma=g/E$.

We have fixed the values of the coupling constants, for the NV$^-$-colour-centres, at $D=2.88$ [GHz] and $E=0.026$ [GHz].

\begin{figure}
\includegraphics[width=6.6cm]{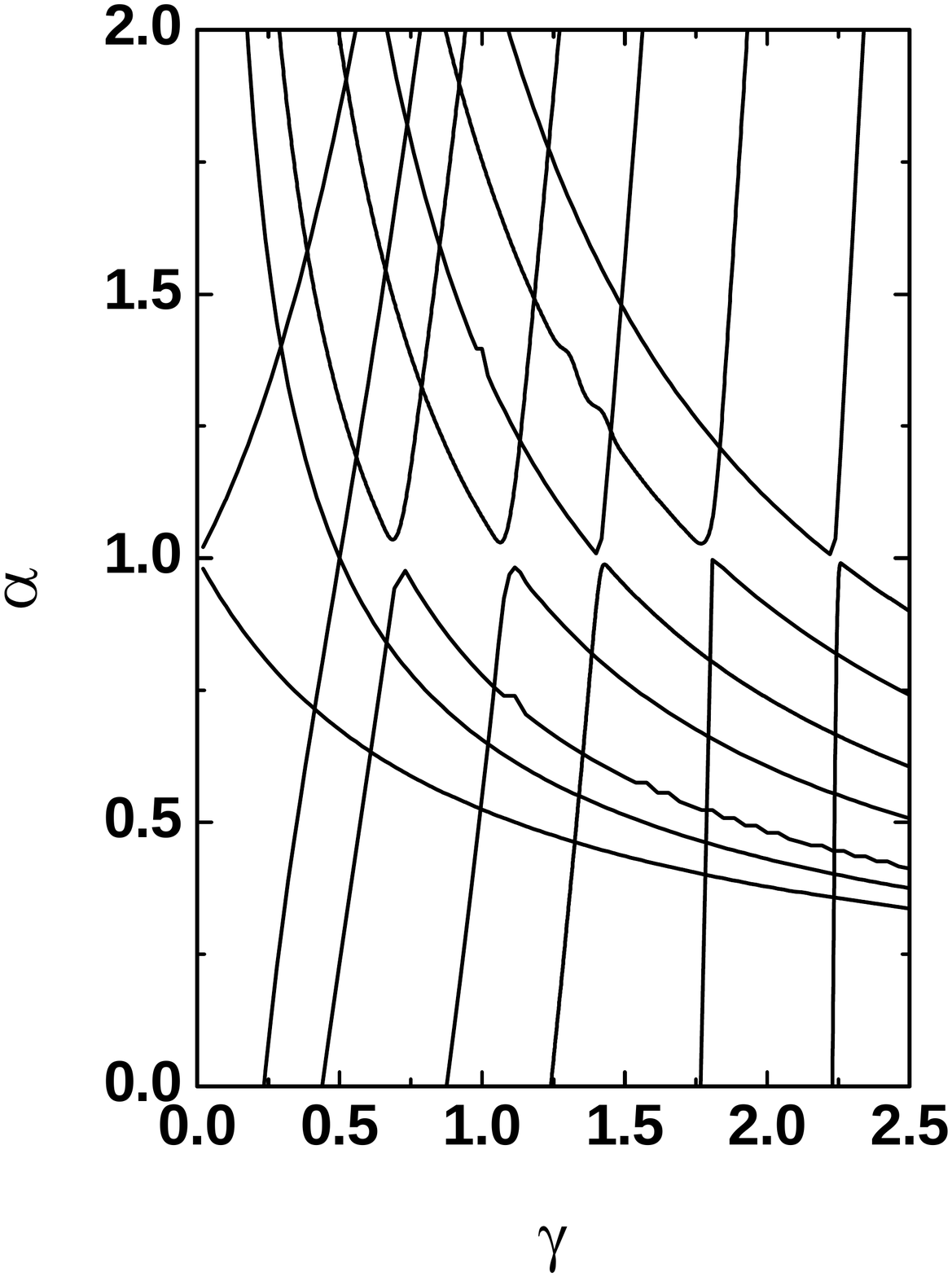}
\caption {Exceptional Points in the plane $(\alpha,~\gamma)$. We have fixed $D=2.88$ [GHz] and $E=0.026$ [GHz]. The Exceptional Points correspond to the coalescense of the eigenvalues $E_{1-}$ and $E_{2-}$. From left to right, the curves correspond to values of $d_-=~0.1,~0.5,~0,~3,~6,~9$, respectively.}\label{fig:fig1}
\end{figure}

In Figure 1, we present the position of EPs in the plane $(\alpha,~\gamma)$. The results correspond to the coalescence of the two
first eigenvalues, $E_{1-}$ and $E_{2-}$. From left to right, the curves correspond to different values of the adimensional parameter $d_-$. We have considered the cases $d_-=~0.1,~0.5,~0,~3,~6,~9$, respectively. We observe a regular pattern for the presence of EPs.

As an example, in Figure 2 and Figure 3, we display the real and the imaginary component of the eigenvalues $E_{1-}$ and $E_{2-}$ as a function of the parameters $\gamma$ and $\alpha$, for the case $d_-=0$, which correspond to fixing $\Delta$, of Eq.(\ref{hami}), to the value $\Delta= 2 D$. As pointed before, there is a region for which both eigenenergies coalescence to the same value, Exceptional Points. At those points the Hamiltonian is non-diagonalizable.
\begin{figure}
\includegraphics[width=6.6cm]{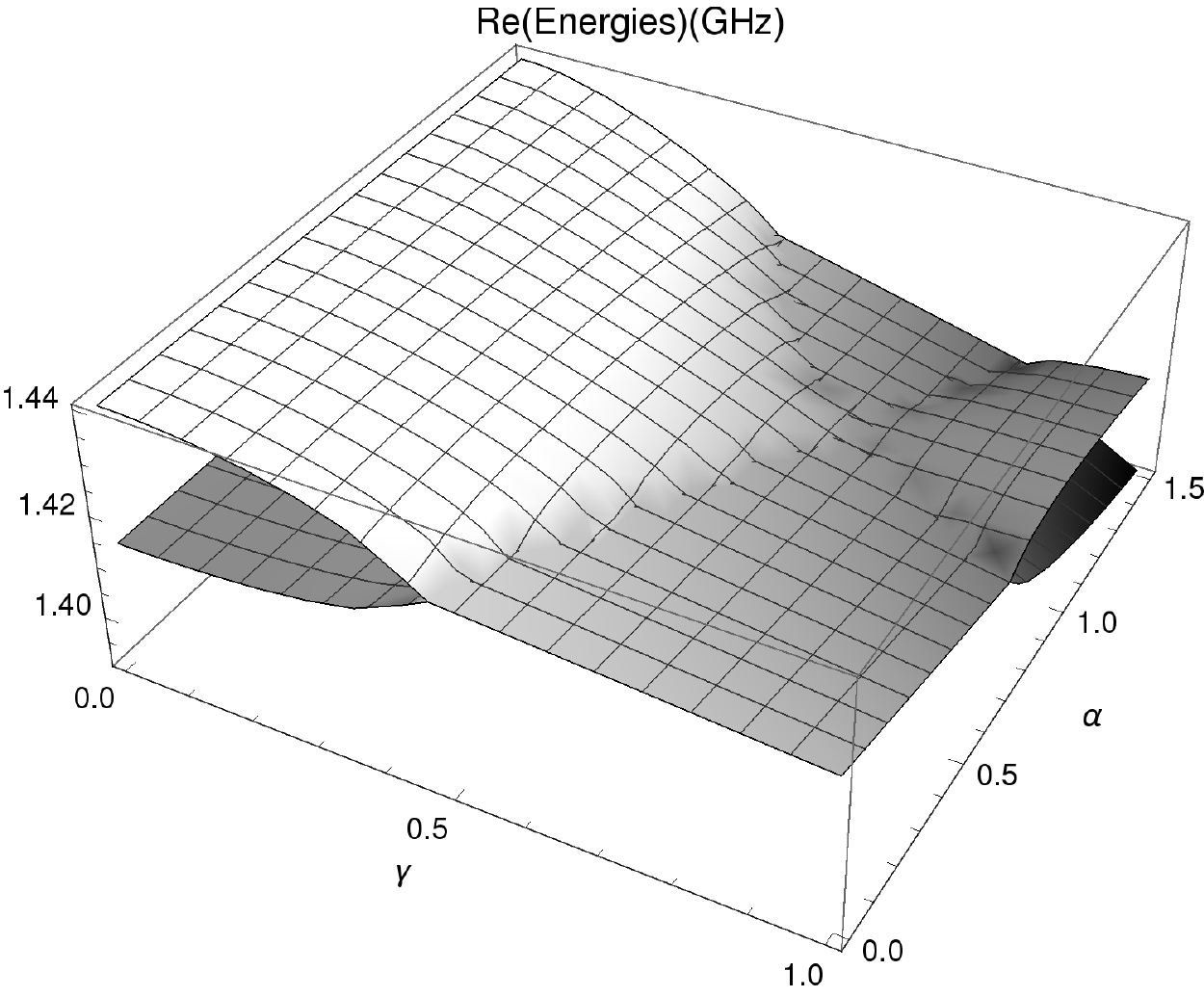}
\caption {Behaviour of the real components of the eigenvalues $E_{1-}$ and $E_{2-}$ as a function of the parameters $\gamma$ and $\alpha$, for $\Delta=2 D$, and $E=0.026$ [GHz].}\label{fig:fig2}
\end{figure}

\begin{figure}
\includegraphics[width=6.6cm]{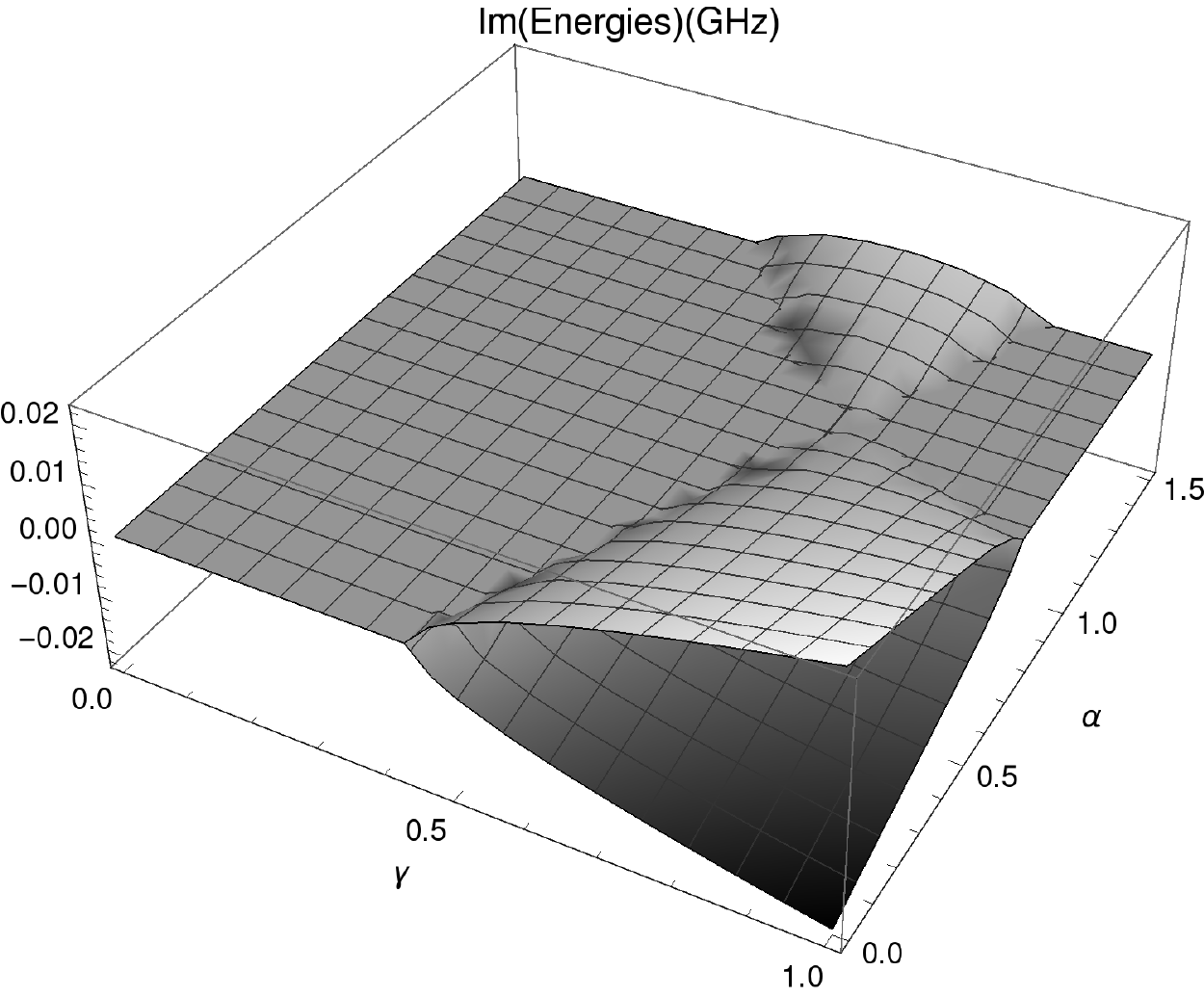}
\caption {Behaviour of the imaginary components of the eigenvalues $E_{1-}$ and $E_{2-}$ as a function of the parameters $\gamma$ and $\alpha$, for the same parameters of Figure 2.}\label{fig:fig3}
\end{figure}

\begin{figure}
\includegraphics[width=6.6cm]{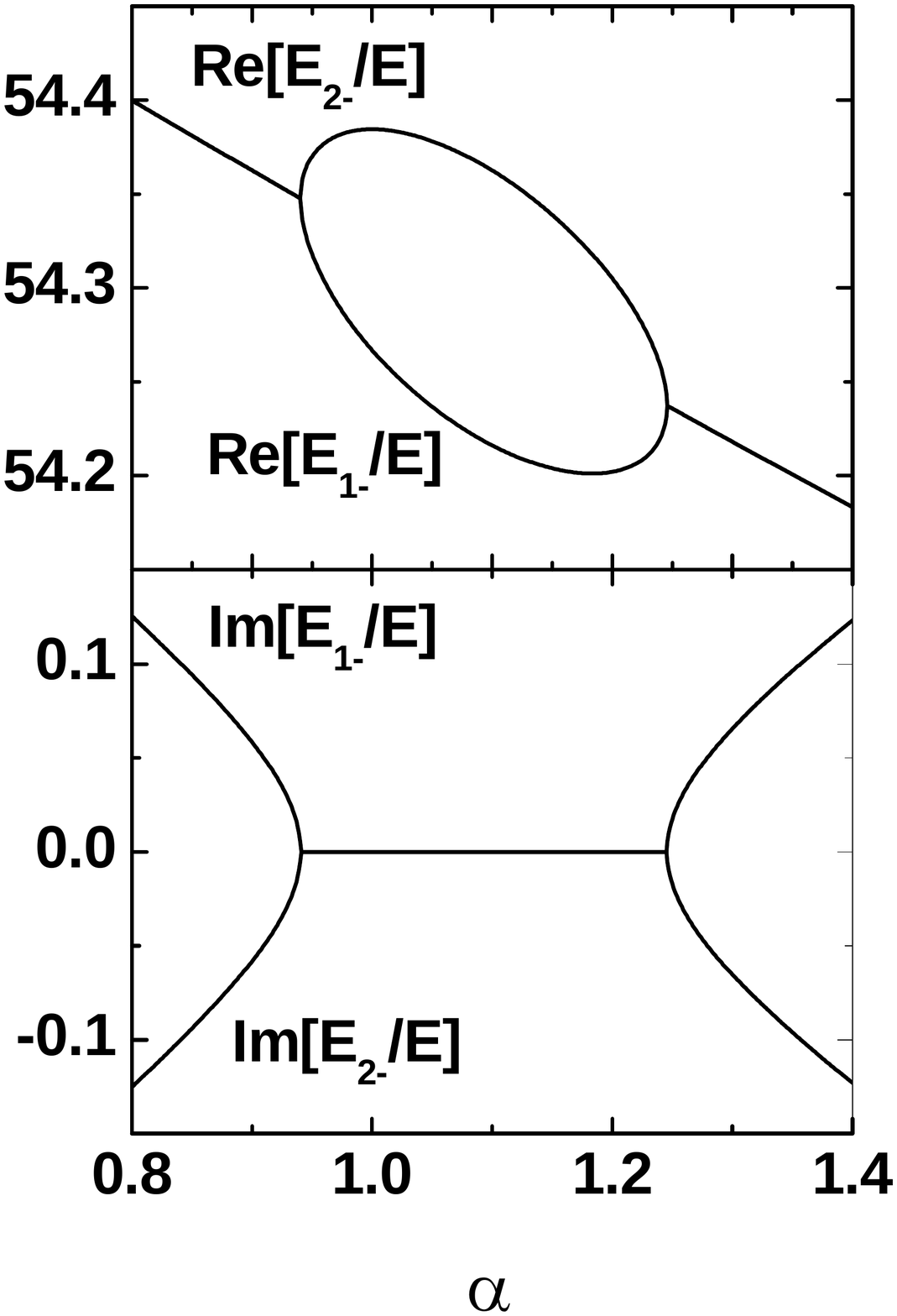}
\caption {Behaviour of the complex eigenvalues $E_{1-}$ and $E_{2-}$, in units of $E$, as a function of the asymmetry parameter $\alpha$, for the parameters adopted in Figures 1 and 2, and for $g=0.02$ [GHz]. }  \label{fig:fig4}
\end{figure}

In what follows, we shall study the time evolution of the system. We shall present the numerical results that we have calculated for a particular value of the coupling interaction constant among the NV$^-$-colour-centres and the SFQ, $g= 0.02$ [GHz]. In Figure 4, we present the behaviour of the complex eigenvalues of the Hamiltonian of Eq. (\ref{hami}), in units of $E$, as a function of the asymmetry parameter $\alpha$. It can be observed two exceptional points, $\alpha \approx 0.94$ and $\alpha \approx 1.24$. Notice that for $\alpha=1$ the Hamiltonian $H$ is a hermitian operator.

The matrix elements  of ${\rm P}$, of Eq.(\ref{transk}), give us information about the transitions probabilities between the vectors of the basis of Eq. (\ref{base}). They determine the evolution in time of the physical observables.
In Figure 5, we display the results we have obtained for the matrix elements of ${\rm P}$, of Eq.(\ref{transk}), as a function of the asymmetry parameter $\alpha$, for the set of parameters of Figure 4. Vertical dotted-lines are drawn at the values of $\alpha$ corresponding to the exceptional points ($\alpha_1 \approx 0.94$ and $\alpha_2 \approx 1.24$). For values of $\alpha$ in the range $(\alpha_1,~ \alpha_2)$ the spectrum is real, as it can be seen from Figure 4. The transition probability curves plotted in this region were computed by making the temporary average in two periods of time. Outside this interval, we have performed the calculation at $t=6$ [$\mu$ sec], when the system has reached the stationary regime. In panels (a) and (b) we show the diagonal elements of ${\rm P}$, while in panels (c) and (d) we present the non-diagonal elements of ${\rm P}$. For $\alpha < 0.94$ and $\alpha>1.24$, the stationary regime is dominated by the transition of the states with positive parity. For values of $\alpha$ corresponding to real spectrum ($ 0.94 < \alpha < 1.24$), all the states have non-zero diagonal entries, though the diagonal elements are dominated by the even states of the basis.
In panel (a) solid-, dashed-,dotted-lines correspond to ${\rm P_{11}(t)}$, ${\rm P_{22}(t)}$, and ${\rm P_{33}(t)}$, respectively. In panel (c) ${\rm P_{12}(t)}$, ${\rm P_{13}(t)}$ and ${\rm P_{23}(t)}$ are presented by solid-, dashed-,dotted-lines, respectively. In panel (b) we plot ${\rm P_{44}(t)}$, ${\rm P_{55}(t)}$ and ${\rm P_{66}(t)}$, while in panel (d) we plot ${\rm P_{ij}(t)}$, for $i,j= 4,~5,~6$.
Notice also that there is a particular value of $\alpha$ for which all non-zero transitions have the same probability, $\alpha \approx 0.824$, we shall show that at this point the steady-state behaves as a Schr\"odinger-cat state.

\begin{figure}
\includegraphics[width=6.6cm]{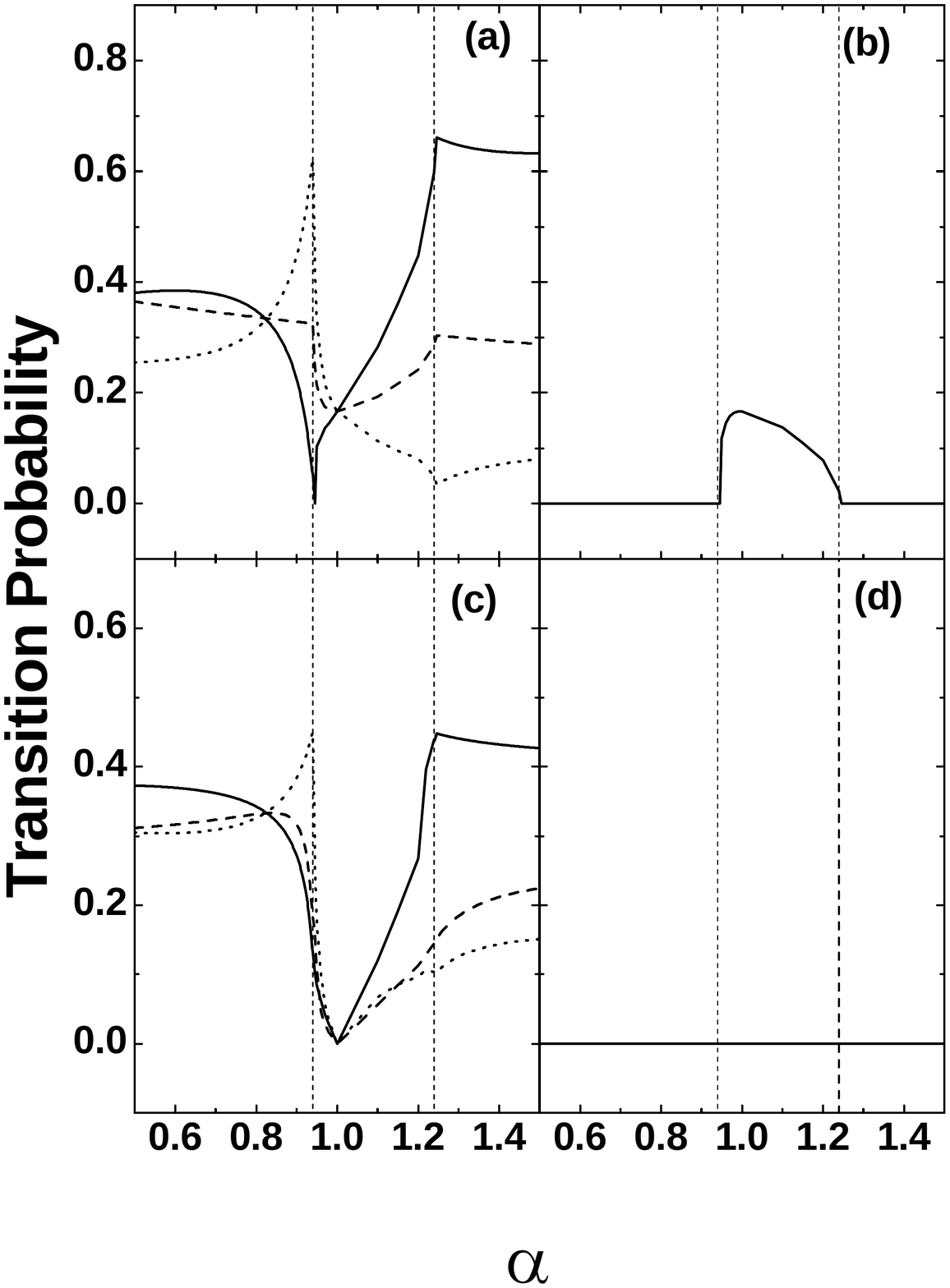}
\caption {Behaviour of the non-zero entries of the matrix ${\rm P(t)}$ of Eq.(\ref{transk}), as a function of the asymmetry parameter $\alpha$, for $g=0.02$ [GHz], $E=0.026$ [GHz], $D=2.88$ [GHz] and $\Delta=2 D$ [GHz]. Vertical dotted-lines are drawn at the values of $\alpha$ corresponding to the exceptional points ($\alpha_1=0.94$ and $\alpha_2 = 1.24$).
For values of $\alpha$ in the range $(0.94,~1.24)$ the transition probability curves, were computed by making the temporary average in two periods of time. For $\alpha < 0.94$ and $\alpha>1.24$ the results presented have been computed at $t=6$ [$\mu$ sec]. In panels (a) and (c) we show the behaviour of the transition probabilities for the even states, while in panels (b) and (d) we show
the transition probalilities for the odd states. In panel (a) solid-, dashed-,dotted-lines correspond to ${\rm P_{11}(t)}$, ${\rm P_{22}(t)}$, and ${\rm P_{33}(t)}$, respectively. In panel (c) ${\rm P_{12}(t)}$, ${\rm P_{13}(t)}$ and ${\rm P_{23}(t)}$ are presented by solid-, dashed-,dotted-lines, respectively. In panel (b) we plot ${\rm P_{44}(t)}$, ${\rm P_{55}(t)}$ and ${\rm P_{66}(t)}$, while in panel (d) we plot ${\rm P_{ij}(t)}$, for $i,j= 4,~5,~6$.}\label{fig:fig5}
\end{figure}

The study of the behaviour of the transition probabilities as a function of time, for different values of $\alpha$,  are presented in Figure 6. In panels (a), (c) and (e) we show the results of the diagonal elements of P(t) as a function of the time, while in panels (b), (d) and (f), the non-diagonal elements of P(t) are displayed. For panels (a) and (b) we have chosen $\alpha=0.6$, for panels (c) and (d) $\alpha=1.1$, and for panels (e) and (f) $\alpha=1.3$, respectively. For values of $\alpha$ for which the spectrum of $H$ has complex-conjugate pair eigenvalues, the matrix elements of P(t) show a non-exponential behaviour, which is the characteristic of systems with gain-loss balance. Meanwhile, for values of $\alpha$ for which the spectrum of $H$ consists only of real eigenenergies, we can observe an oscillatory patron.

\begin{figure}
\includegraphics[width=6.6cm]{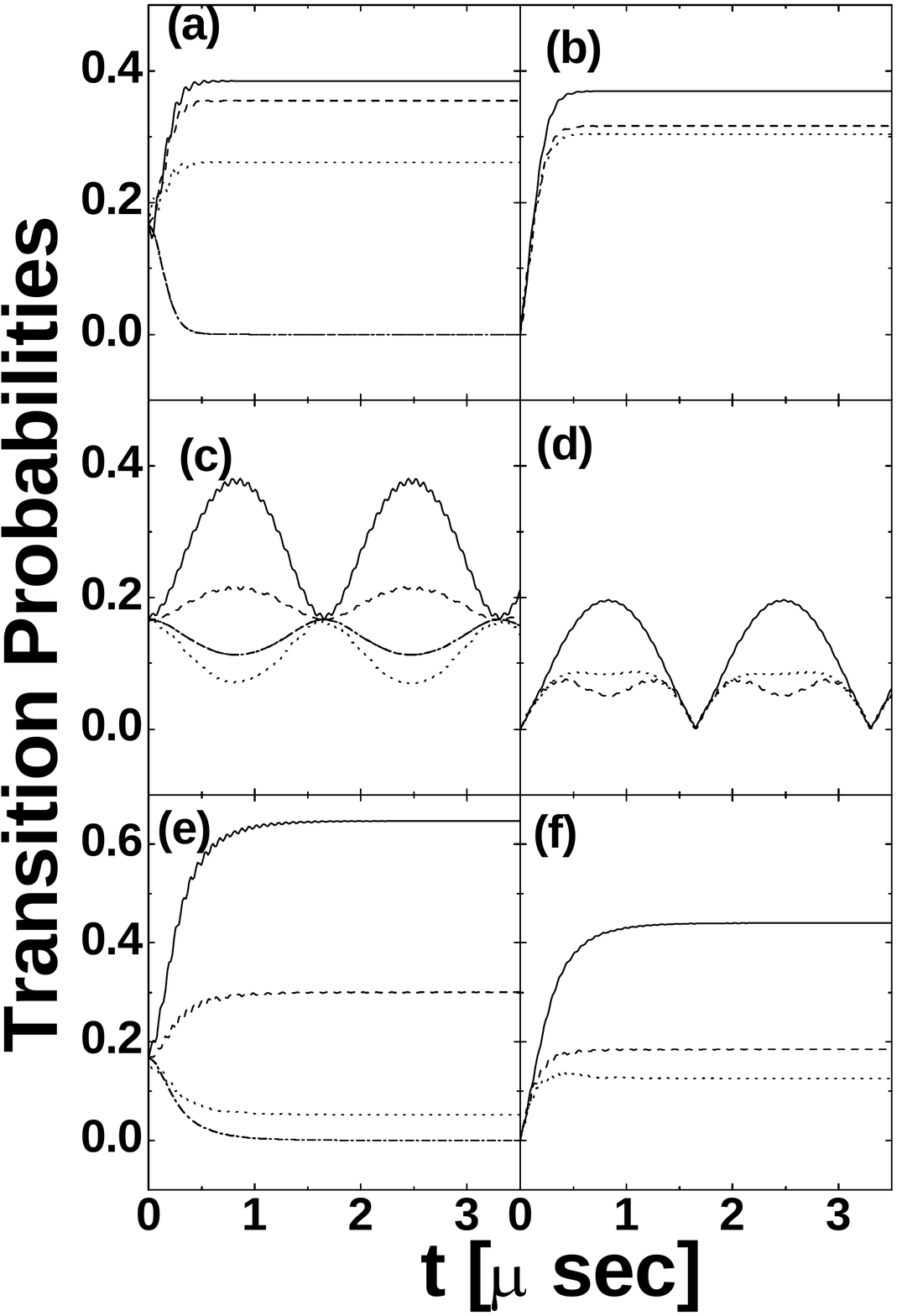}
\caption {Behaviour of the non-zero entries of the matrix ${\rm P(t)}$ of Eq.(\ref{transk}), as a function of the time, for the parameters of Figure 5. The results presented have been computed at $\alpha=0.6$ (Insests (a) and (b)), $\alpha=1.1$ (Insests (c) and (d)), and $\alpha=1.3$ (Insests (e) and (f)), respectively. In Insests (a), (c) and (e), the behaviour of the diagonal elements of P(t) are shown. Solid-, dashed-,dotted-lines correspond to ${\rm P_{11}(t)}$, ${\rm P_{22}(t)}$, and ${\rm P_{33}(t)}$, respectively.
${\rm P_{44}(t)}$, ${\rm P_{55}(t)}$ and ${\rm P_{66}(t)}$ are plotted with dashed-solid lines. In panels (b), (d) and (f), the non-diagonal entries of P(t) are shown. ${\rm P_{12}(t)}$, ${\rm P_{13}(t)}$ and ${\rm P_{23}(t)}$ are presented by solid-, dashed-,dotted-lines, respectively.}\label{fig:fig6}
\end{figure}
\begin{figure}
\includegraphics[width=8cm]{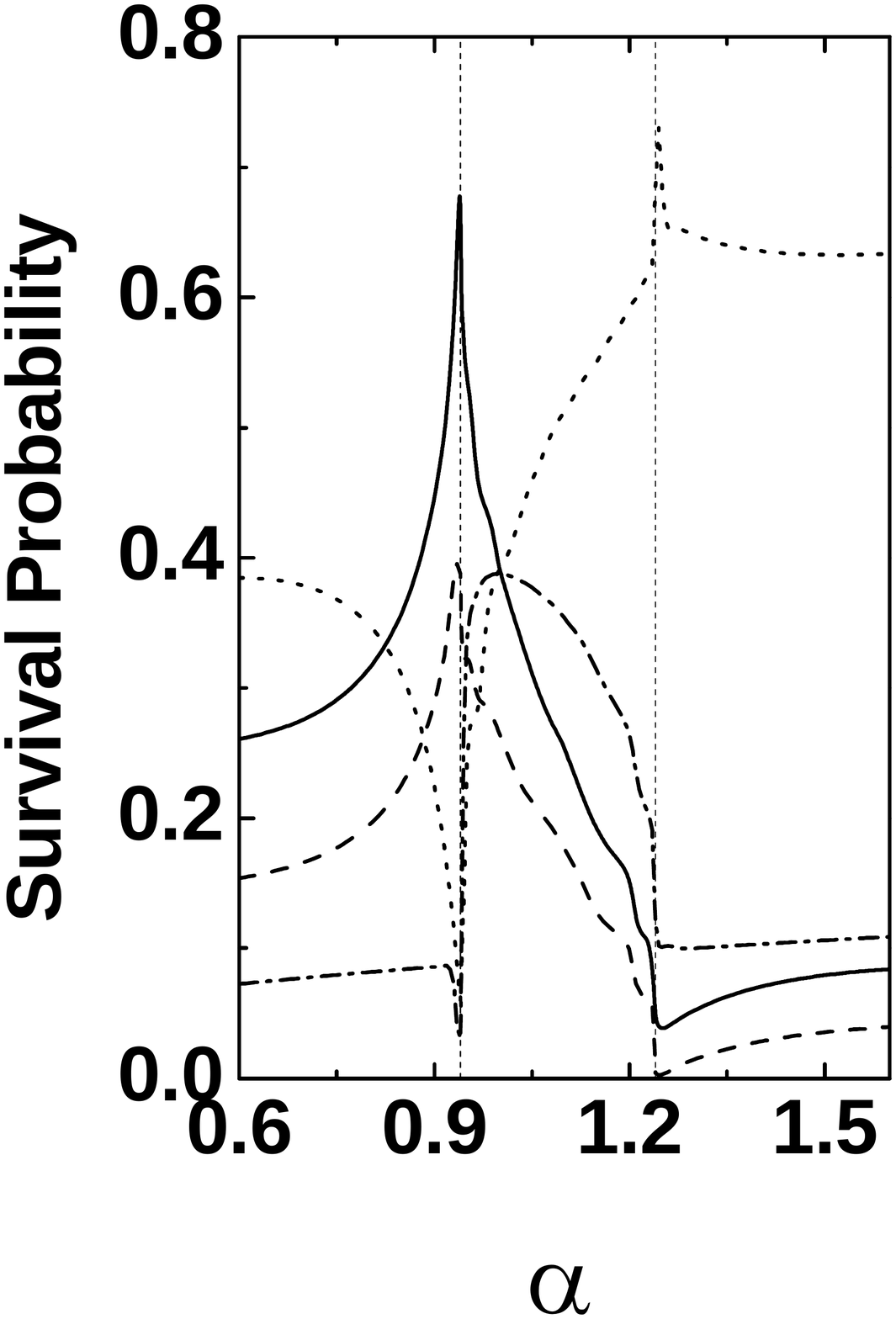}
\caption { Survival Probability, p(t) of Eq.(\ref{prob}), as a function of $\alpha$. The parameters are those of the previous Figures. Initially, the superconducting qubits are prepared in its ground state, and the NVs in a coherent state with $(\theta_0, \phi_0)=(0,0)$ (solid line), $(\pi/4,0)$ (dashed line), $(\pi/2,0)$ (dashed-dotted-line), and $(\pi,0)$ (dotted line), respectively. For values of $\alpha$ in the range $(0.94,~1.24)$ we plot the average value in two periods of time. For $\alpha < 0.94$ and $\alpha>1.24$ the results presented have been computed at $t=6$ [$\mu$ sec].}\label{fig:fig7}
\end{figure}

\begin{figure}
\includegraphics[width=8cm]{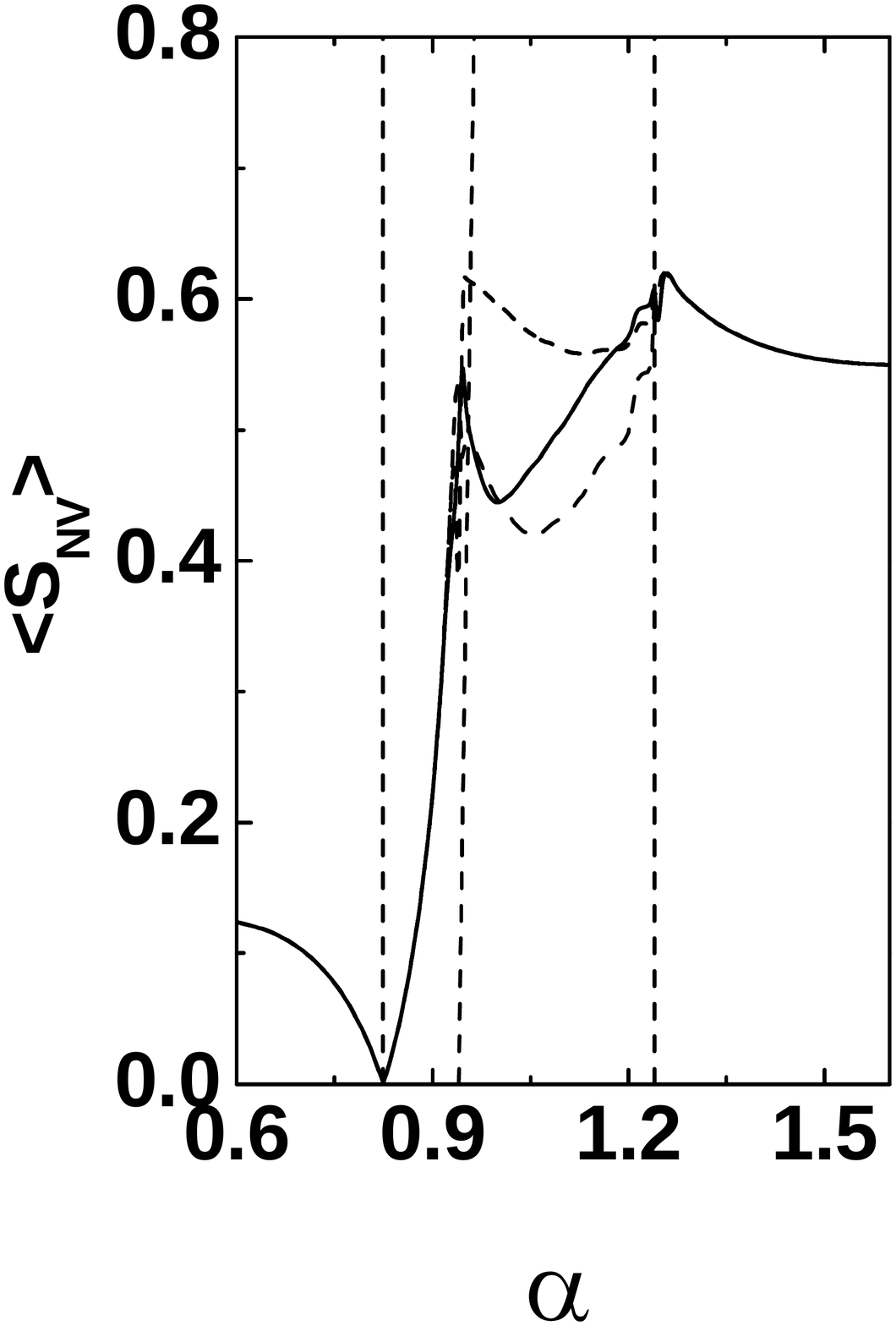}
\caption { Mean Value of the Spin of the NV$^-$-colour-centres, $\langle {\bf S}_{NV} \rangle$, as a function of $\alpha$. The parameters are those of the previous Figures. Initially, the superconducting qubits are prepared in its ground state, and the NVs in a coherent state $(\theta_0, \phi_0)=(0,0)$ (solid line), $(\pi/4,0)$ (dashed line), $(\pi/2,0)$ (dashed-dotted-line), and $(\pi,0)$ (dotted line), respectively. For values of $\alpha$ in the range $(0.94,~1.24)$ we plot the average value in two periods of time. For $\alpha < 0.94$ and $\alpha>1.24$ the results presented have been computed at $t=6$ [$\mu$ sec].}\label{fig:fig8}
\end{figure}

\begin{figure}
\includegraphics[width=8cm]{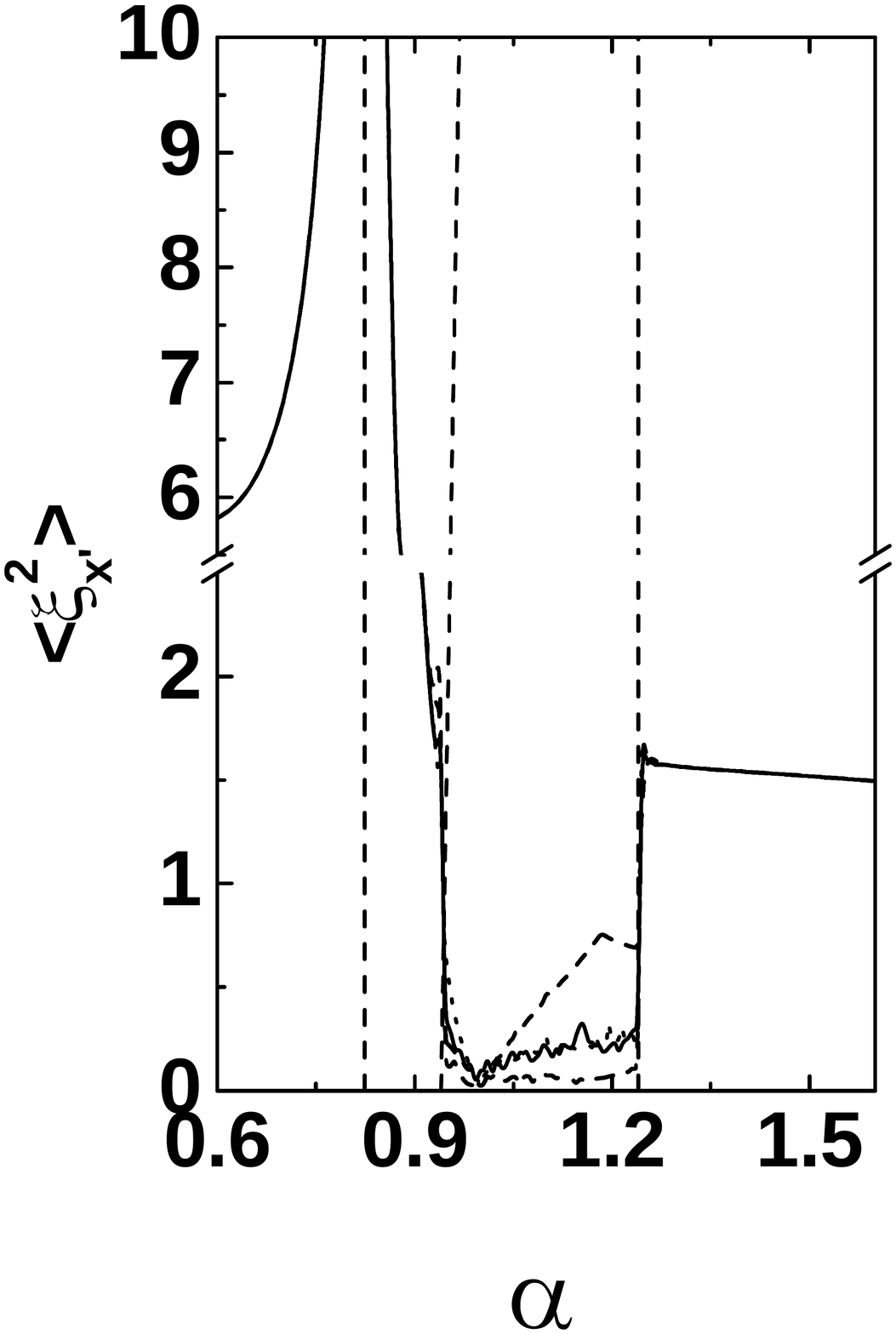}
\caption { Squeezing Parameter, $\xi^2_{NV}$, for the NV$^-$-colour-centres, as a function of $\alpha$. The parameters are those of the previous Figures. Initially, the superconducting qubits are prepared in its ground state, and the NVs in a coherent state $(\theta_0, \phi_0)=(0,0)$ (solid line), $(\pi/4,0)$ (dashed line), $(\pi/2,0)$ (dashed-dotted-line), and $(\pi,0)$ (dotted line), respectively. For values of $\alpha$ in the range $(0.94,~1.24)$ we plot the minimun value of  $\xi^2_{NV}$ in two periods. For $\alpha < 0.94$ and $\alpha>1.24$ the results presented have been computed at $t=6$ [$\mu$ sec].}\label{fig:fig9}
\end{figure}

The results we have obtained for the Survival Probability, ${\rm p(t)}$ of Eq.(\ref{prob}), as a function of $\alpha$, are displayed in Figure 7.  The parameters are those of the previous Figures. Initially, the SFQ is prepared in its ground state, and the NVs in a coherent state with $(\theta_0, \phi_0)=(0,0)$, solid line, $(\pi/4,0)$, dashed line, $(\pi/2,0)$, dashed-dotted-line, and $(\pi,0)$, dotted line, respectively. As in the previous Figure, for values of $\alpha$ in the range $(0.94,~ 1.24)$ the curves were computed by making the temporary average in two periods of time. For $\alpha < 0.94$ and $\alpha>1.24$ the results presented have been computed at $t=6$ [$\mu$ sec], stationary regime. From the Figure, it can be inferred that the Survival Probability depends both on the initial state and on the asymmetry parameter.
For $(\theta_0, \phi_0)=(0,0)$ and for $(\pi/4,0)$, it is observed an increase in the survival probability at the first EP, while for $(\theta_0, \phi_0)=(\pi,0)$, the increase in the survival probability is observed at the second EP.

In Figure 8 the Mean Value of the total spin operator of the NV$^-$-colour-centres, $\langle {\bf S}_{NV} \rangle$, is plotted as a function of $\alpha$. The parameters are those of the previous Figures. Initially, the superconducting qubit is prepared in its ground state, and the NVs in a coherent state with $(\theta_0, \phi_0)=(0,0)$, solid line, $(\pi/4,0)$, dashed line, $(\pi/2,0)$, dashed-dotted-line, and $(\pi,0)$, dotted line, respectively. As in the previous Figure, for values of $\alpha$ in the range $(0.94,~ 1.24)$ the curves were computed by making the temporary average in two periods of time. For $\alpha < 0.94$ and $\alpha>1.24$ the results presented have been computed at $t=6$ [$\mu$ sec]. For $\alpha < 0.94$
and for $\alpha >1.24$, the Mean Value of the Spin operator of the NV$^-$-colour-centres is independent of the preparation of the initial state. It should be noticed the existence of a value of $\alpha$ for which $\langle {\bf S}_{NV}\rangle=0$, $\alpha=0.82402$.

The Squeezing Parameter for the NV$^-$-colour-centres, $\xi^2_{NV}$, as a function of $\alpha$ is presented in Figure 9. The parameters are those of the previous Figures. Initially, the superconducting qubits are prepared in its ground state, and the NVs in a coherent state with $(\theta_0, \phi_0)=(0,0)$, solid line, $(\pi/4,0)$, dashed line, $(\pi/2,0)$, dashed-dotted-line, and $(\pi,0)$, dotted line, respectively. For values of $\alpha$ in the range $(0.94,~1.24)$ we plot the minimum value of  $\xi^2_{NV}$ in two periods of time. For $\alpha < 0.94$ and $\alpha>1.24$ the results presented have been computed at $t=6$ [$\mu$ sec]. This Figure is in agreement with the previous one. At the value of $\alpha=0.82402$, for which $\langle {\bf S}_{NV} \rangle=0$, the value of the squeezing parameter diverges, the system is maximally anti-squeezed. In the region of real spectrum, the system shows an oscillatory patron, with the minimum values shown in the curve.

We shall discuss in what follows the behaviour of the Squeezing parameter, the mean value of the total spin and the survival probability, as a function of the time, for different values of the asymmetry parameter, $\alpha$.

\begin{figure}
\includegraphics[width=8cm]{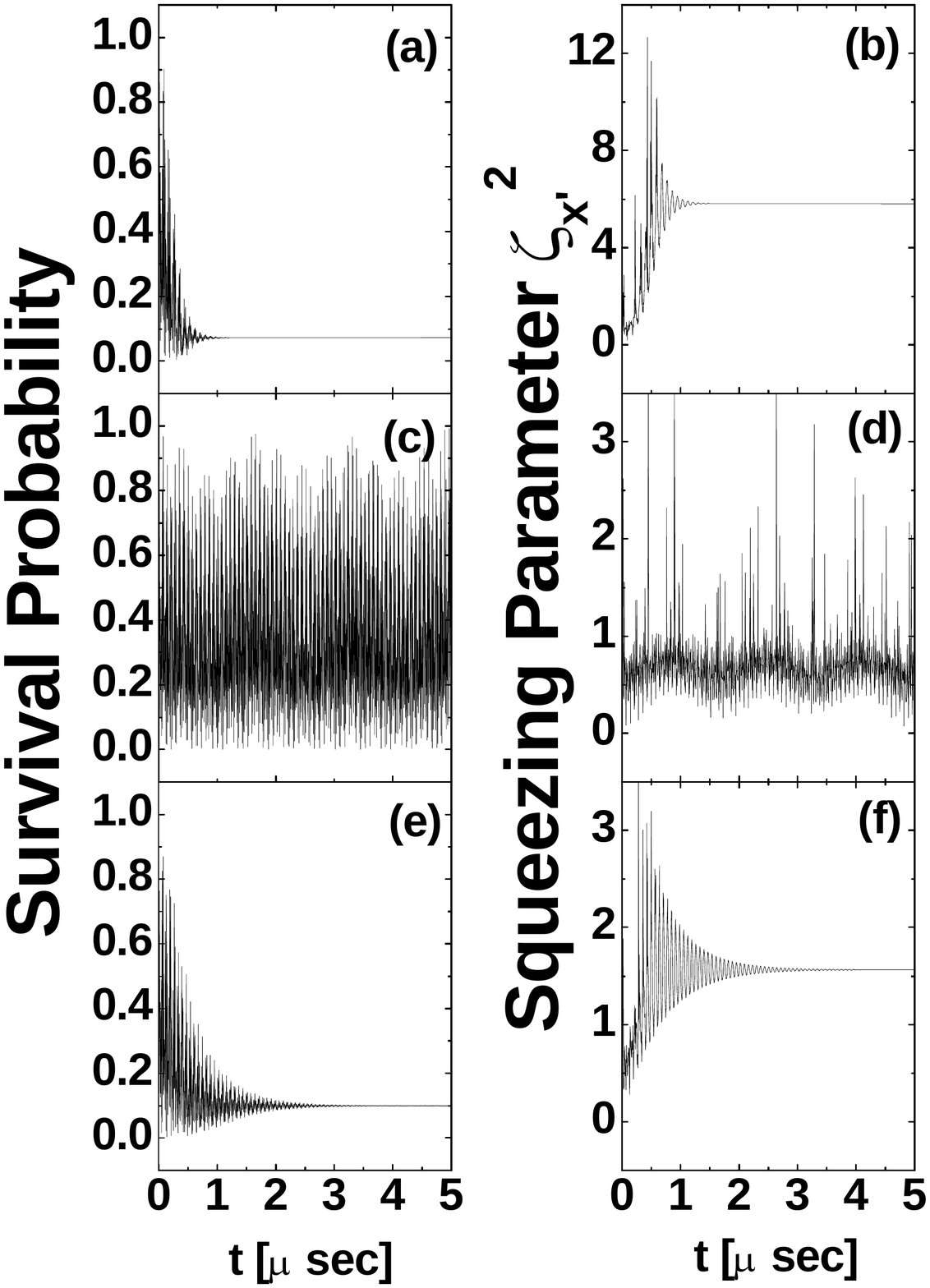}
\caption {Squeezing Parameter and Survival Probability, for the NV ensemble, as a function of the time. The parameters are those of the previous Figures. Initially, the superconducting qubits are prepared in its ground state, and the NVs in a coherent state with $(\theta_0, \phi_0)=(\pi/2,0)$. In panels (a), (c) and (e), we show the behaviour of the Survival Probability, ${\rm p(t)}$, for $\alpha=0.6,~1.1$ and $1.3$, respectively. In panels (b), (d) and (e), we show the behaviour of the Squeezing Parameter, $\xi^2_{NV}$, for $\alpha=0.6,~1.1$ and $1.3$, respectively. }\label{fig:fig10}
\end{figure}

\begin{figure}
\includegraphics[width=8cm]{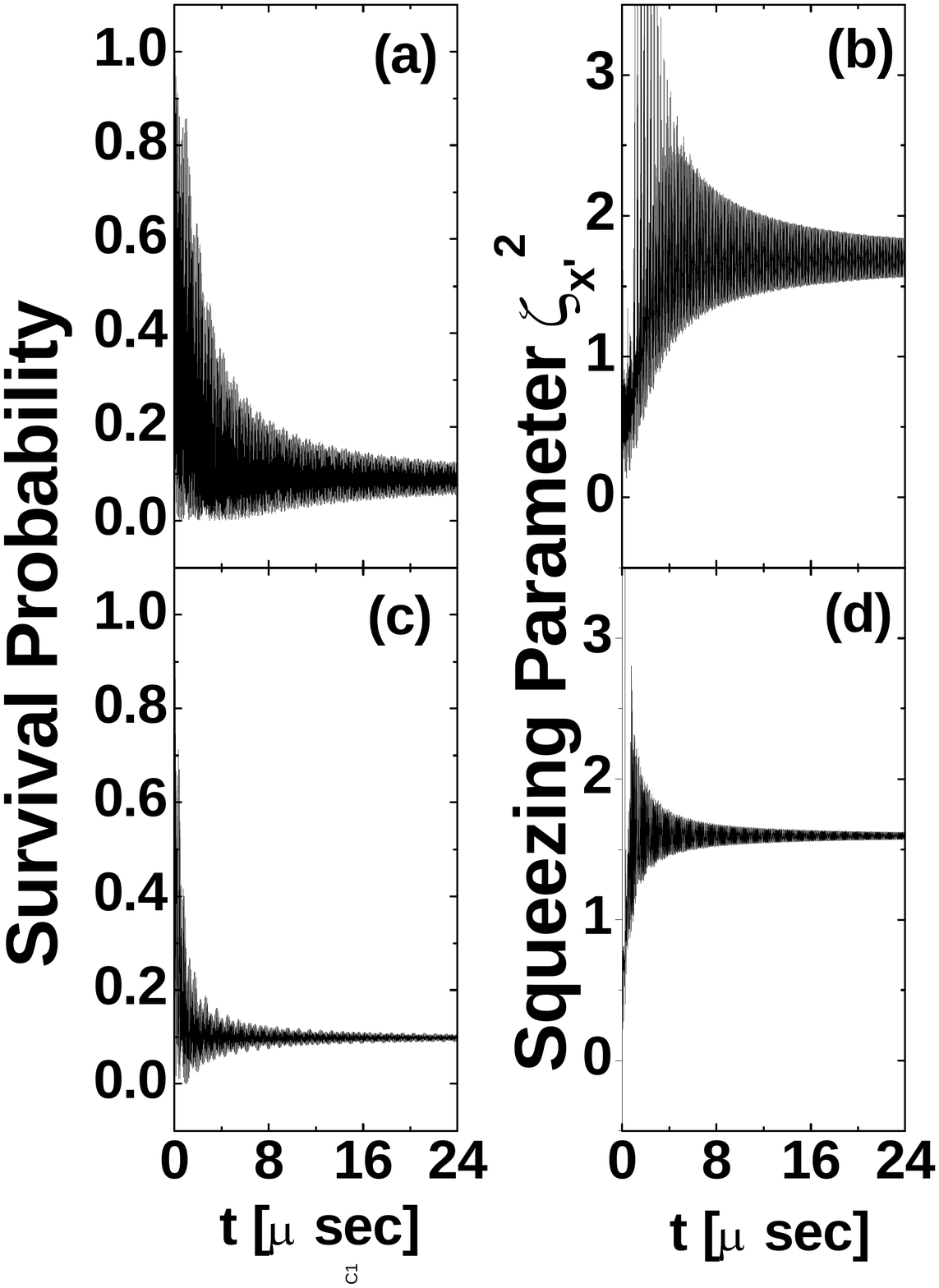}
\caption {Squeezing Parameter and Survival Probability, for the NV ensemble, as a function of the time. The parameters are those of the previous Figures. Initially, the superconducting qubits are prepared in its ground state, and the NVs in a coherent state with $(\theta_0, \phi_0)=(\pi/2,0)$. In panels (a) and (c), we show the behaviour of the Survival Probability, p(t), for $\alpha=0~0.9424$ and $1.24556$, respectively.
 In panels (b)-(d), we show the behaviour of the Squeezing Parameter, $\xi^2_{NV}$, for $\alpha=~0.9424$ and $1.24556$, respectively. }\label{fig:fig11}
\end{figure}

In Figure 10 and 11, we present the results we have obtained for the Squeezing Parameter, $\xi^2_{NV}$ of Eq.(\ref{sqx}), and for the Survival Probability, ${\rm p(t)}$ of Eq.(\ref{prob}), as a function of the time. The parameters are those of the previous Figures. Initially, the superconducting qubits are prepared in its ground state, and the NVs in a coherent state with $(\theta_0, \phi_0)=(\pi/2,0)$. The curves of Figure 10 were computed at $\alpha=0.6,~1.1$ and $1.3$, while the curves of Figure 11 were computed at the values of the asymmetry parameter $\alpha$ corresponding to the two EPs, $\alpha=~0.9424$ and $1.24556$.
From the Figures, it can be observed that at the early stages of the time the state evolves as a squeezed state. This pattern is preserved at values of $\alpha$ corresponding to the regime of real spectrum, see Figure 9-(d). For other values of the asymmetry parameter, the steady-state is an anti-squeezed state.
Concerning the Survival Probability, ${\rm p(t)}$, the results shown in Figures 10 and 11, can be understood in terms of the behaviour of the matrix elements of P(t), which have been presented in Figure 5. The states with the largest probability, are the state $| k_{qb}, k_{NV} \rangle= |0,2\rangle$ and $|0,0\rangle$, which are present in the initial state of Figure 10 and 11. Consequently, we can use the information of Figure 5 to prepare initial states robust against decoherence. The results of Figure 11 support the idea that at the EPs the initial state evolves into a non-trivial state as reported in \cite{scexp}. In both Figures, it can be observed that the Survival Probability does not obey an exponential decay law.

\subsection{Schr\"odinger-cat states.}

We shall discuss in some detail the case of $\alpha=0.82402$. As it can be observed in Figure 4, all the probabilities $P_{ij}$, in the stationary regime, are equally distributed. From Figure 8 it can be observed that initial state of the system evolves to a steady-state with mean value of the total spin of the NVs equals zero, $\langle {\bf S}_{NV}\rangle=0$, independent of the initial state chosen. This is a sign of the presence of Schr\"odinger-cat states.

In Figure 12, panels (a), (c) and (e), we show the results obtained for the Mean Value of the Total Spin of the ensemble of NVs of colour centres, $\langle {\bf S} _{NV} \rangle$, as a function of the time. The Squeezing Parameter, $\xi^2_x (t)$, for a given Initial State as a function of the time is displayed in panels (b), (d) and (f). We have considered different initial states. We have adopted an Initial ground state for the superconducting qubit. We assume that the NV$^-$-colour-centres are prepared in a coherent state, with $(\theta_0, \phi_0)=(0,0)$ for panels (a) and (b), $(\pi/2,0)$ for panels (c) and (d), and $(\pi,0)$ for panels (e) and (f), respectively. From the Figure, it can be observed that initial state of the system evolves into a steady-state with mean value of the total spin of the NVs equals zero, $\langle {\bf S}_{NV}\rangle=0$, independent of the initial state chosen.
At short times, the spin dynamics is caused by the spin-spin interaction of the ensemble of NV$^-$-colour-centres, namely the OAT term of $H$. The $H_{OAT}$ of $H$ is responsible for the squeezing pattern observed. At longer times, the growth of spin correlations causes both the depolarization observed in panels (a)-(c), and the increase of the squeezing parameter \cite{antisq}.

In Figures 13 to 15, we plot the Discrete SU(2) Wigner function for the NV$^-$-colour-centres \cite{wigner1,wigner2,wigner3,noswig1,noswig2}. In the left panels, we display the behaviour of $W(\theta,\phi)$ for the initial state and in the right ones for the steady-state. Initially, the superconducting qubits are prepared in its ground state, and the NVs in a coherent state with $(\theta_0, \phi_0)=(\pi/2,0)$, $(\pi/4,0)$ and $(0,0)$, Figure 13, 14 and 15, respectively. To obtain the Discrete SU(2) Wigner function for the NVs, we have computed the reduced density matrix for both subsystems. It can be observed that the steady-state is independent of the preparation of the inial state. The SU(2) Wigner function is composed by two Wigner functions, namely one corresponding to the superposition of two coherent states, $ |I(\pi/2,0) \rangle + | I(-\pi/2,0)\rangle$, and a pure state, $S_+|0 \rangle$. Thus, the steady-state is a Scr\"odinger cat-like state.

\begin{figure}
\includegraphics[width=8cm]{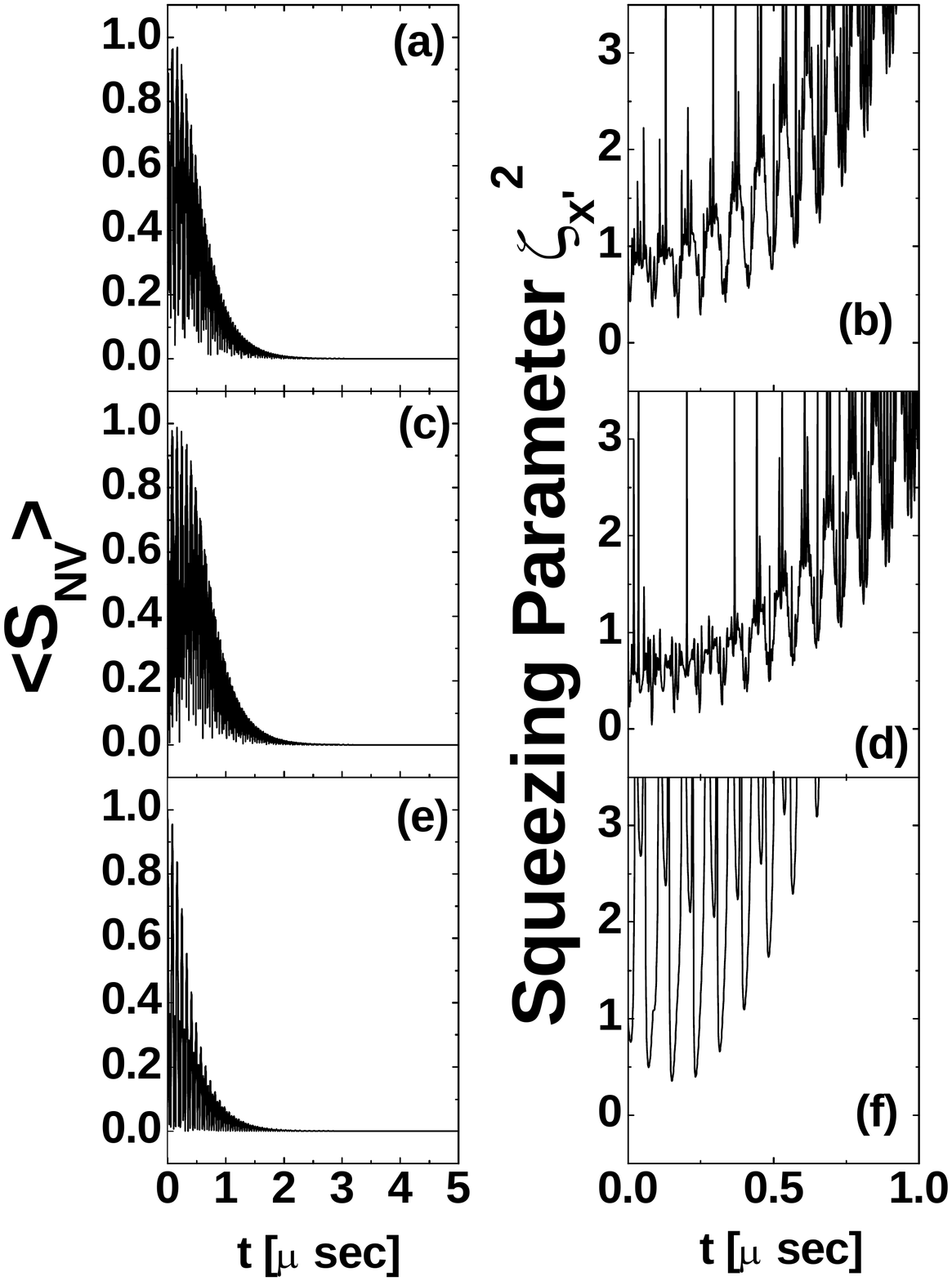}
\caption {The Mean Value of the Total Spin of the ensemble of NV$^-$-colour-centres, $\langle {\bf S} _{NV} \rangle$, as a function of the time, is displayed in panels (a), (c) and (e). The Squeezing Parameter, $\xi^2_x (t)$, for a given Initial State, as a function of the time, is displayed in panels (b), (d) and (f). We have considered different initial states. We have adopted an Initial ground state for the superconducting qubits. We assume that the NV$^-$-colour-centres are prepared in a coherent state, with $(\theta_0, \phi_0)=(0,0)$ for panels (a) and (b), $(\pi/2,0)$ for panels (c) and (d), and $(\pi,0)$ for panels (e) and (f), respectively.}\label{fig:fig12}
\end{figure}

\begin{figure}
\includegraphics[width=10cm]{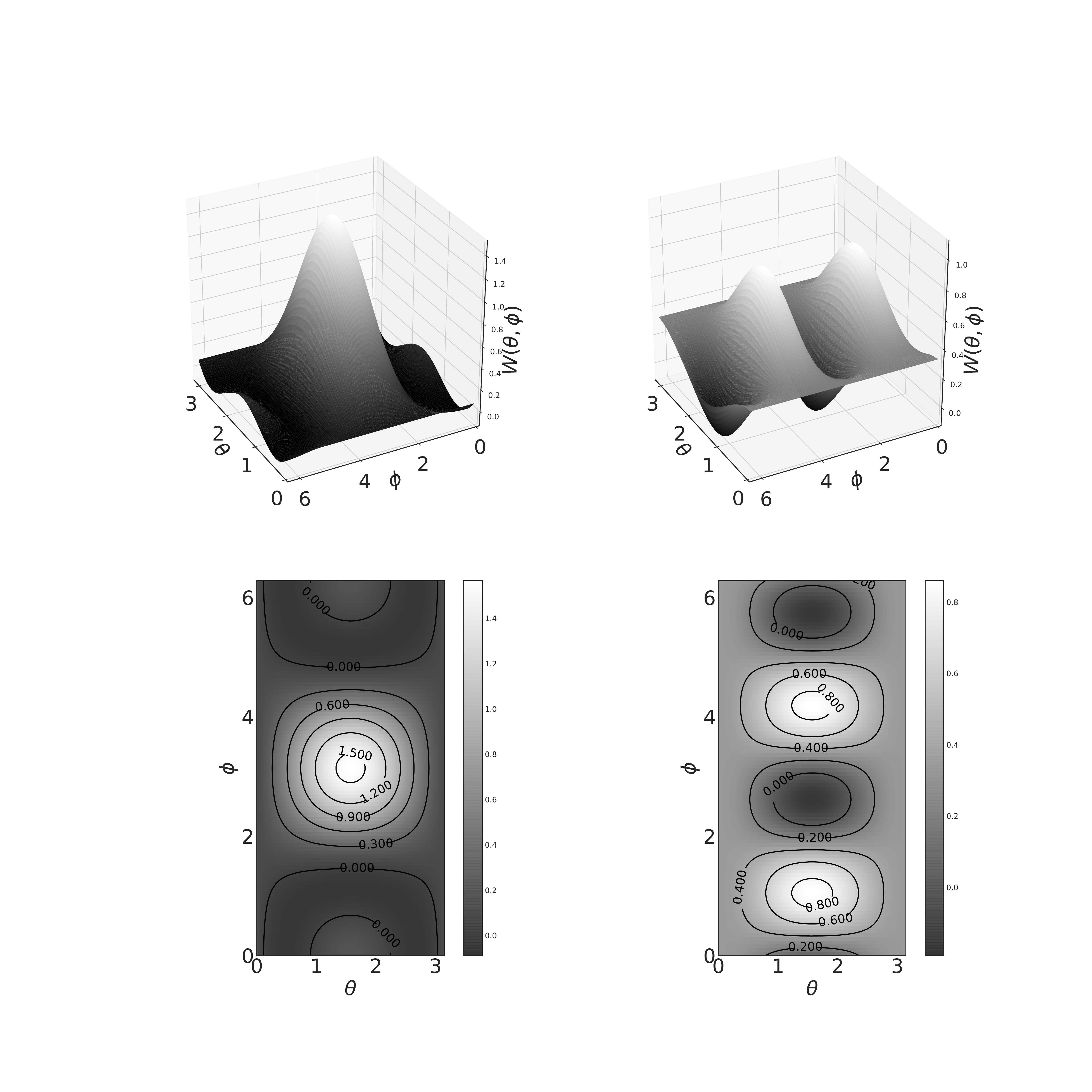}
\caption {Discrete Wigner Function $W(\theta,\phi)$ of the NV$^-$-colour-centres. In the left panels we display the behaviour of $W(\theta,\phi)$ for the Initial State and in the left ones in the steady-state. Initially, the superconducting qubits are prepared in its ground state, and the NVs in a coherent state with $(\theta_0, \phi_0)=(\pi/2,0)$.}\label{fig:fig13}
\end{figure}

\begin{figure}
\includegraphics[width=10cm]{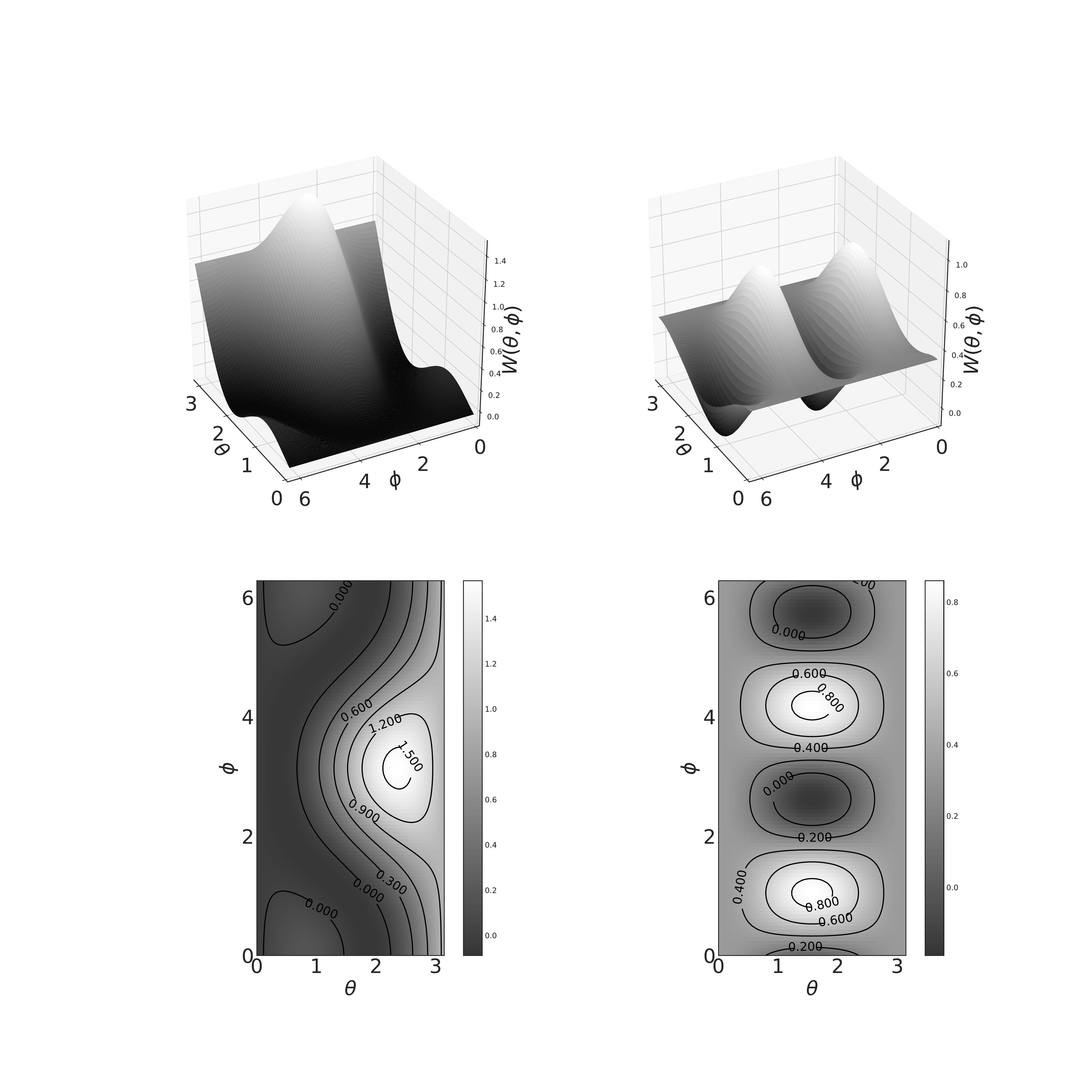}
\caption {Discrete Wigner Function $W(\theta,\phi)$ of the NV$^-$-colour-centres. In the left panels we display the behaviour of $W(\theta,\phi)$ for the Initial State and in the left ones in the steady-state. Initially, the superconducting qubits are prepared in its ground state, and the NVs in a coherent state with $(\theta_0, \phi_0)=(\pi/4,0)$.}\label{fig:fig14}
\end{figure}

\begin{figure}
\includegraphics[width=10cm]{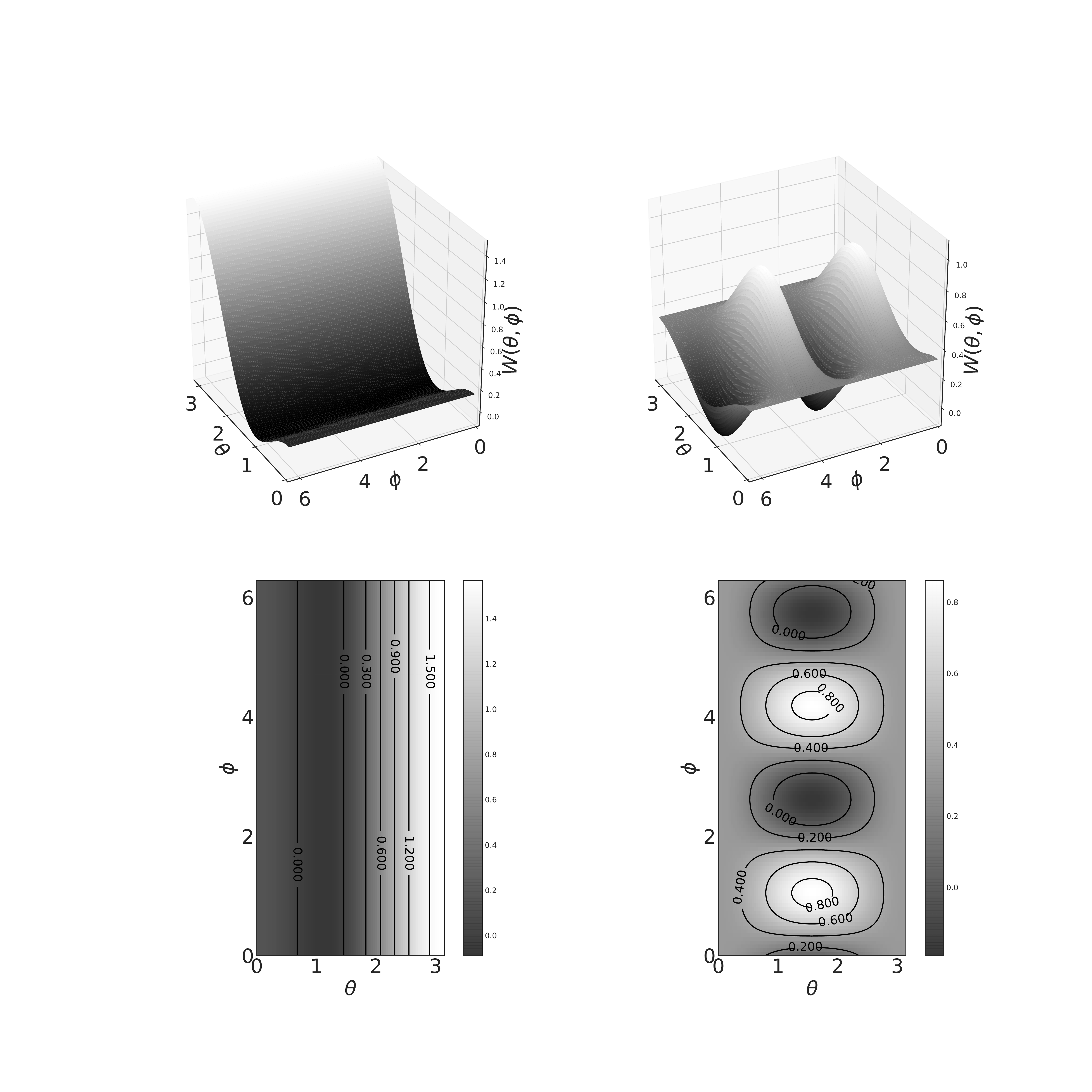}
\caption {Discrete Wigner Function $W(\theta,\phi)$ of the NV$^-$-colour-centres. In the left panels we display the behaviour of $W(\theta,\phi)$ for the Initial State and in the left ones in the steady-state. Initially, the superconducting qubits are prepared in its ground state, and the NVs in a coherent state with $(\theta_0, \phi_0)=(0,0)$.}\label{fig:fig15}
\end{figure}

In Figure 16, we present the points in the plane $(\alpha,~\gamma)$, for which the steady-state behaves as a Schr\"odinger-cat states,  for the case $d_-=~0$.

\begin{figure}
\includegraphics[width=6.6cm]{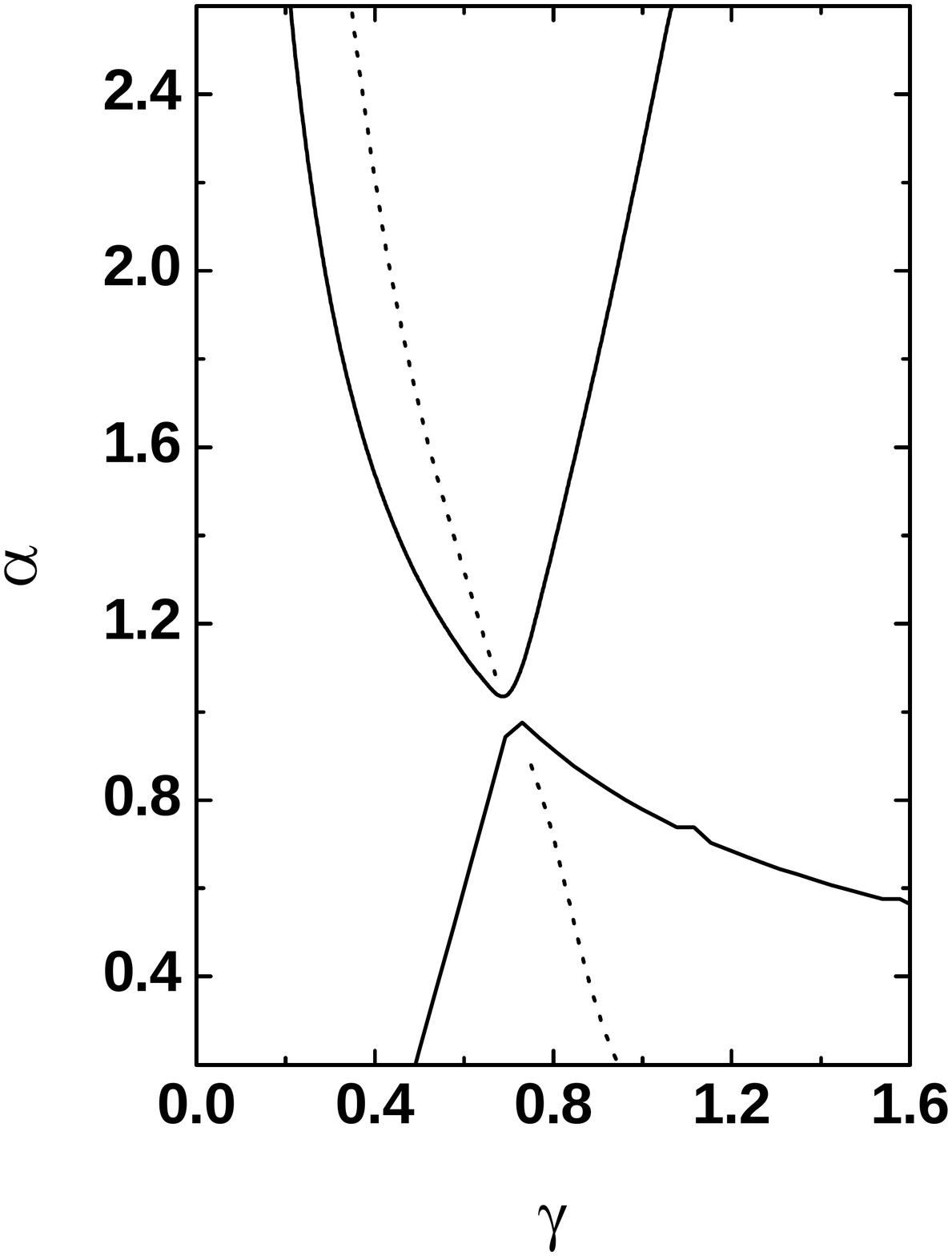}
\caption {Points in the plane $(\alpha,~\gamma)$, at which steady Schr\"odinger-cat states occur, are displayed in dashed -line.  We have considered the case $d_-=0$, with $D=2.88$ [GHz] and $E=0.026$ [GHz]. The solid-line corrresponds to the contour plot of Exceptional Points.}\label{fig:fig16}
\end{figure}

\begin{figure}
\includegraphics[width=6.6cm]{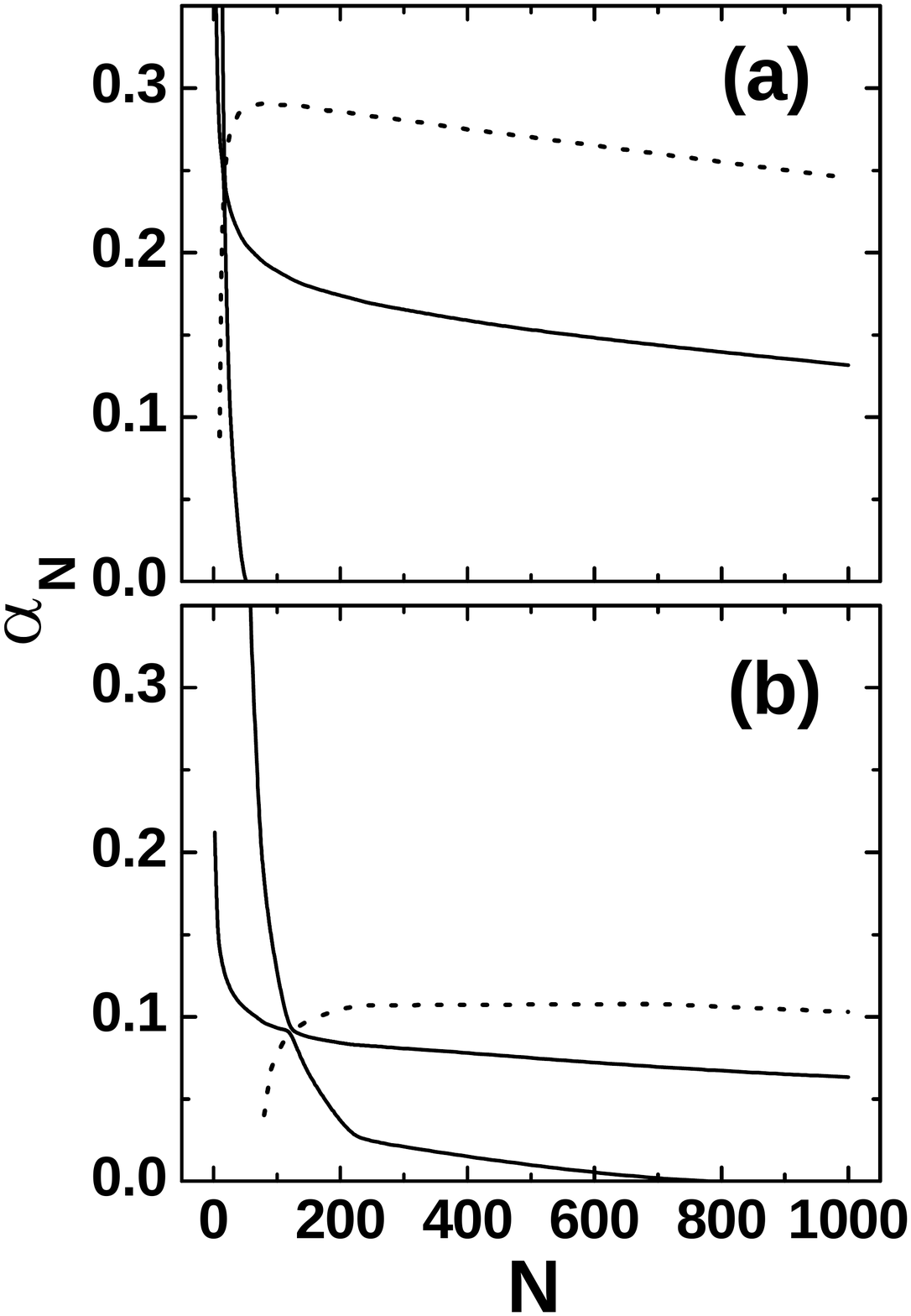}
\caption {Points in the plane $\left (\alpha_N,~\gamma \right)$, $\alpha_N=\alpha/\sqrt{N}$, at which steady Schr\"odinger-cat states occur, are displayed in dashed-line. The solid-line corrresponds to the contour plot of Exceptional Points.  We have considered the case $d_-=0$, with $D=2.88$ [GHz] and $E=0.026$ [GHz]. In panel (a) and (b) we plot results for $g=0.02$ [GeV] ($\gamma_N=1/(13 \sqrt{\rm N})$) and $g=0.052$ [GeV] ($\gamma_N=2/\sqrt{\rm N} $), respectively.}\label{fig:fig17}
\end{figure}

\subsection{Hybrid system with $N$ NV$^-$-colour-centres in diamond.}

To study qualitatively the behaviour of the system when the number of NV$^-$-colour-centres in diamond is increased, we can perform a Holstein-Primakoff boson mapping \cite{klein} of the Hamiltonian of Eq.(\ref{h0}). We can write

\beqn
S_+=b_+ \sqrt{N- b_+b_-}, ~ S_-=S_+^\dagger, ~ S_z=b_+b_--\frac N 2,\nonumber \\
\label{mapbos}
\eeqn
with $[b,b^\dagger]=1$. To leading order in the boson mapping of Eq.(\ref{mapbos}), the Hamiltonian reads

\beqn
H_B&=&\frac {E_{qb}} {2} \sigma_z +D \left ( b_+b_--\frac N 2 \right)^2+\frac {N E}{2} (b_+^2+b_-^2)+ \nonumber \\
& & \frac {\sqrt{N} g}{4}\left(~\cos \alpha ~\sigma_{z } + \sin \alpha  ~\sigma_{x } \right) \left (b_+ + \alpha b_- \right).
\label{hbos}
\eeqn
The nonlinearity introduced by the square-root term in Eq.(\ref{mapbos})
ensures that two excitations can not take place at the same
spin. If we consider delocalized spin waves involving a large
number of spins compared to the number of excitations, the
probability that a given spin is excited is inversely proportional
to the number of spins N. Therefore, as long as only a few
delocalized spin excitations are considered, it is reasonable to
adopt the approximation, $S_+ \approx b_+ \sqrt{N}$ \cite{molmer}.

From the form $H_B$, it can be inferred that the adimensional parameter $\gamma$ must be scaled as $\gamma \rightarrow \gamma_N= \gamma/ \sqrt{N}$.

We shall take into account the previous re-scaling of the coupling constants to analyse the exact results, which have been computed from the Hamiltonian of Eq. (\ref{h0}).
In Figures 17, we present the contour plots for the appearance of EPs and of steady Schr\"odinger-cat 
states in the plane $\left ( \alpha_N,N \right)$, with $\alpha_N=\alpha/\sqrt{{\rm N}}$. The solid-line corrresponds to the contour plot of EPs, while contour plot for steady Schr\"odinger-cat 
states are displayed in dashed-line. We have considered the case $d_-=0$, with $D=2.88$ [GHz] and $E=0.026$ [GHz]. 
In panel (a) and (b) we plot results for $g=0.02$ [GeV] ($\gamma_N=1/(13 \sqrt{\rm N})$) and $g=0.052$ [GeV] ($\gamma_N=2/\sqrt{\rm N} $), respectively.
Though we plot continuous lines in Figure 17, they are meant to guide the eyes. They are valid for the case of even number of particles. For the present set of parameters, and in the range $0 \le \alpha_N \le 1$, we have not observed EPs in the case of an odd number of NVs (we have checked that for other sets of parameters, $d_-\neq 0$ and larger values of $E$, EPs are present for an odd number of particles, i.e. for N$=3$ we have observed the presence of EPs at values of $\Delta=D/4$ [GHz], $E=1.8$ [GHz], $-2.0 \le g \le 2.0$ [GHz] and $0.4 \le \alpha \le 1.5$). From the curves, it can be observed the presence of a regular pattern of Exceptional Points and a regular pattern for steady Schr\"odinger-cat states.

\section{Conclusions}\label{conclusions}

In this work, we have studied the time evolution of a hybrid system consisting of NV$^-$-colour-centres in diamond in interaction with a superconducting flux qubit. We have modelled the dynamics of the system through a non-hermitian Hamiltonian, to take into account the effect of the environment. Though the Hamiltonian does not preserve ${\cal PT}$-symmetry, the spectrum consists of real eigenvalues or complex-conjugate pair eigenvalues, and it shows the characteristics features of a system with gain-loss balance \cite{gainloss}. We observed a regular pattern of Exceptional Points, as a function of the parameters of the model. At these points, the initial state evolves into a non-trivial steady-state. The study of the matrix elements of the Fourier Transform of the Green Matrix provides information on the transition probabilities of the states of the original base as a function of time. Thus, we can prepare robust initial states by combining the states of the base of Eq.(\ref{base}) which show large transition probabilities at long intervals of time. The Survival Probability can be analysed to account for this effect. At Exceptional Points, the Survival Probability increases considerably, depending on the initial state adopted. It is observed that in the regime of real spectrum, the state evolves in time showing a periodical patron of collapses and revivals. In this regime, the states are periodically squeezed. While in the regime of complex-conjugate pair spectrum the steady-state is not a squeezed state, and anti-squeezing is observed.
The initial state of the system evolves to a steady-state with mean value of the total spin of the NVs equals zero, $\langle {\bf S}_{NV}\rangle=0$, independent of the initial state chosen. This a sign of the presence of Schr\"odinger-cat states. That is, the steady-state behaves as a Schrodiger-cat state, it can be written as the superposition of two coherent spin states.
At short times, the spin dynamics is caused by the spin-spin interaction of the ensemble of NV$^-$-colour-centres, namely the OAT term of $H$. The $H_{OAT}$ of $H$ is responsible for the squeezing pattern observed. At longer times, the growth of spin correlations causes both the depolarization observed and the increase of the squeezing parameter. We have extended the previous analysis to systems with a larger number of NV$^-$-colour-centres in diamond. We have rescaled the adimensional constant $\gamma$ as $\gamma/ \sqrt{N}$, and we have found a regular pattern of Exceptional Points and a regular pattern of steady Schr\"odinger-cat states in the plane $(\alpha_N,{\rm N})$.

Work is in progress concerning the analysis of hybrid systems with more than one superconducting flux qubit, interacting with an ensemble of NV$^-$-colour-centres in diamond \cite{nosap17}.

\begin{acknowledgments}
This work was partially supported by the National Research Council
of Argentine (PIP 282, CONICET) and by the Agencia Nacional de
Promocion Cientifica (PICT 001103, ANPCYT) of Argentina.
\end{acknowledgments}

\end{document}